\documentclass[twocolumn]{aastex62}

\graphicspath{{./}{figures/}}

\submitjournal{AJ}

\shorttitle{Evolved blue stragglers in LMC clusters}
\shortauthors{Li et al.}

\begin{document}

\title{Blue straggler stars beyond the Milky Way. III. Detection of
  evolved blue straggler candidates in Large Magellanic Cloud clusters}

\correspondingauthor{Chengyuan Li}
\email{chengyuan.li@mq.edu.au}

\author{Chengyuan Li} 
\affil{Department of Physics and Astronomy, Macquarie University,
  Sydney, NSW 2109, Australia}

\author{Licai Deng}
\affiliation{Key Laboratory for Optical Astronomy, National
  Astronomical Observatories, Chinese Academy of Sciences, 20A Datun
  Road}

\author{Kenji Bekki}
\affiliation{International Centre for Radio Astronomy Research/The
  University of Western Australia, M468, 35 Stirling Highway, Crawley,
  WA6009, Australia}

\author{Jongsuk Hong}
\affiliation{Kavli Institute for Astronomy and Astrophysics, Peking
  University, Yi He Yuan Lu 5, Hai Dian District, Beijing 100871,
  China}

\author{Richard de Grijs}
\affiliation{Department of Physics and Astronomy, Macquarie
  University, Sydney, NSW 2109, Australia}
\affiliation{International Space Science Institute--Beijing, 1
  Nanertiao, Zhongguancun, Hai Dian District, Beijing 100190, China}
\affiliation{Kavli Institute for Astronomy and Astrophysics, Peking
  University, Yi He Yuan Lu 5, Hai Dian District, Beijing 100871,
  China}

\author{Bi-Qing For}
\affiliation{International Centre for Radio Astronomy Research/The
  University of Western Australia, M468, 35 Stirling Highway, Crawley,
  WA6009, Australia}
\affiliation{Australian Research Council, Centre of Excellence for
  All-sky Astrophysics in 3 Dimensions}


\begin{abstract}
We analyze {\sl Hubble Space Telescope} observations of nine Large
Magellanic Cloud star clusters with ages of 1--2 Gyr to search for
evolved counterparts of blue straggler stars. Near the red clump
regions in the clusters' color--magnitude diagrams, we find branches
of evolved stars that are much brighter than normal evolved stars. We
examine the effects of photometric artifacts, differential reddening,
and field contamination. We conclude that these bright evolved stars
cannot be explained by any of these effects. Our statistical
  tests show that the contributions of photometric uncertainties and
crowding effects, as well as that owing to differential reddening, to
these bright evolved stars are insufficient to fully explain the
  presence of these bright evolved stars. Based on isochrone fitting,
  we also ruled out the possibility that these bright evolved stars
  could be reproduced by an internal chemical abundance spread. The
spatial distributions of the bright evolved stars exhibit clear
concentrations that cannot be explained by homogeneously distributed
field stars. This is further confirmed based on Monte Carlo-based
tests. By comparing our observations with stellar evolution models, we
find that the masses of most of bright evolved stars do not exceed
twice the average mass of normal evolved stars. We suggest that these
bright evolved stars are, in fact, evolved blue straggler stars.
\end{abstract}

\keywords{blue stragglers --- galaxies: star clusters ---
  Hertzsprung-Russell and C-M diagrams --- Magellanic Clouds}

\section{Introduction} \label{S1}

Blue straggler stars (BSSs), which are commonly found in dense stellar
systems like globular clusters (GCs) or open clusters (OPs), are
  stars located along the blue extension of the main-sequence turnoff
  (MSTO) regions in the color--magnitude diagrmas (CMDs) of star
  clusters. It is thought that they are main-sequence (MS)-like stars
that are significantly more massive than the cluster's bulk population
\cite[e.g.,][]{Ferr12a,Bald16a,Raso17a}. They represent a population
of rejuvenated stars formed through dynamical evolution of old
population stars. Since the host clusters have already exhausted all
of their gas, BSSs are unlikely traditional star-forming products,
presumed to have resulted from the collapse of a molecular cloud. Two
leading mechanisms invoked to retain BSSs on the MS include binary
mass transfer and the eventual merger of both binary components
\citep{Andr06a,Hill76a}, and direct stellar collisions
\citep{Mccr64a}.

The evolved counterparts of most BSSs (eBSSs) in old GCs are
photometrically indistinguishable from normal stars undergoing
advanced stages of stellar evolution. The best locus to search for
eBSSs in old OCs is the horizontal branch (HB), where stars are placed
according to their mass. Using this approach, \cite{Ferr97a} detected
19 eBSSs in the core region of the GC M3. Possible eBSSs can also be
identified by studying the number of stars in the asymptotic giant
branch (AGB) region. \cite{Becc06a} found a significant excess of AGB
stars in the GC 47 Tuc. They interpreted this pattern as a result of
the progeny of massive stars originating from the evolution of binary
systems. Using stellar spectra and chemical abundances, \cite{Ferr16a}
identified an eBSSs mass that is significantly higher than the MSTO
mass in the GC 47 Tuc.

In old stellar populations like those found in most GCs, stars that
have evolved into red giants now supported by helium fusion in their
cores will develop a red clump (RC) in the CMD. The RC properties are
independent of their internal age spread, and thus they end up with
the same luminosity. For younger stellar populations ($\lesssim$2
Gyr), such as those dominating young massive clusters (YMCs) or OCs,
the magnitude extent of the RCs is a function of age. Inclusion of an
evolved blue straggler population will contribute to the extension of
the RC, thus making eBSSs distinguishable from normal RC
stars. Studying younger star clusters is therefore a promising way to
photometrically search for eBSSs. However, almost all young Galactic
clusters are located in the Galactic disk. They are affected by severe
foreground extinction, which will elongate the RCs in the CMDs, thus
masking the difference between eBSSs and normal RC stars. Star
clusters in the Large Magellanic Cloud (LMC) cover a much more
extended age range than those in the Milky Way (MW). In addition, the
LMC is located at high Galactic latitude, where Galactic extinction is
small. LMC clusters are therefore ideal targets to search for eBSSs in
younger star clusters.

Only few studies have explored BSSs in Magellanic Cloud star
clusters. \cite{Li13a} examined 162 BSSs in the old LMC GC Hodge
11. They found that its blue straggler population is split into two
clumps characterized by different colors. Those in the cluster's
central region are systematically bluer than their counterparts
further out. \cite{Li18a} found two distinct BSS populations in a
young (1--2 Gyr-old) LMC GC, NGC 2173, a situation that is similar to
those found in GGCs \citep[e.g.,][]{Ferr09a,Dale13a,Simu14a}. However,
unlike GGCs exhibiting bifurcated BSS populations, no evidence of any
putative post-core-collapse event was detected in this
cluster. \cite{Li18b} analyzed the BSSs in the young LMC GC NGC
2213. They found that although this cluster's population of BSSs has a
half-mass relaxation time that is shorter than the cluster's
isochronal age, the BSSs in NGC 2213 are not fully segregated. They
suggest that this is likely caused by interactions between the BSSs'
progenitor binaries and black hole subsystems, as well as by dynamical
disruption of binaries in this young GC. \cite{Sun18a} studied
  BSSs in 24 Magellanic Cloud clusters. They derived a sub-linear
  correlation between the number of BSSs in the cluster cores and the
  clusters' core masses. They concluded that this may be an indication
  of a binary origin for these BSSs, which is consistent with similar
  conclusions regarding BSSs in Milky Way GCs \citep{Knig09a}.

In this paper, we search for eBSSs in nine LMC young GCs. We find that
all of these clusters harbor samples of evolved stars that are located
in the bright extension of their RCs. We rule out the possibilities
that these bright evolved stars are simply caused by photometric
artifacts (such as crowding) and measurement uncertainties, by
differential reddening, by field-star contamination, or by an
  internal chemical abundance spread. For almost all these bright
evolved stars, their masses do not exceed twice the average mass of
normal stars if we assume that they are all single stellar systems. We
suggest that the most straightforward interpretation of our
observations is that they are eBSSs.

This article is organized as follows. In Section \ref{S2} we introduce
the details of our observations and the data reduction. Section
\ref{S3} presents our main results. In Section \ref{S4} we provide a
brief physical discussion about the origins of these bright evolved
stars. Section \ref{S5} contains a summary.

\section{Observations and Data Reduction} \label{S2}

All clusters studied in this work were observed with the {\sl HST}'s
Ultraviolet and Visible channel of the Wide Field Camera 3 (UVIS/WFC3)
or with the Wide Field Channel of the Advanced Camera for Surveys
(ACS/WFC). For each cluster, except for NGC 1644, we also adopted a
parallel observation of a nearby region as reference field, which will
be used to statistically estimate the level of field-star
contamination. For NGC 1644 we did not find a proper parallel
observation that can represent its reference field, but we will show
that the edge of the science image that includes the main cluster is
sufficient to represent a reference field. Relevant information
pertaining to the data set is present in Table \ref{T1}.

\begin{table*}
  \begin{center}
\caption{Description of the science images used in this article.}\label{T1}
  \begin{tabular}{l l l l l l l}\hline
    Cluster      &  Camera	  & Exposure time & Filter & Program ID & PI name \\\hline
    NGC 1644$^*$ 	& ACS/WFC & 250 s	 & F555W	& GO-9891 & G. Gilmore & \\
    			&  		    & 170 s	 & F814W &  		  & 		       & \\\hline
    NGC 1651 (Cluster)	& UVIS/WFC3  & 120 s+600 s+720 s & F475W & GO-12257 & L. Girardi & \\
    			&			&  30 s + 2$\times$700 s & F814W &	     &		       & \\
    NGC 1651 (Ref. Field)  & ACS/WFC & 2$\times$500 s & F475W & GO-12257 & L. Girardi & \\
    					& 		    & 2$\times$500 s & F814W & 		 & 		 & \\\hline
    NGC 1783 (Cluster)	& ACS/WFC & 40 s + 2$\times$340 s & F555W	& GO-10595	& P. Goudfrooij & \\
			&			&  8 s + 2$\times$340 s  & F814W &	    		 &		       & \\
    NGC 1783 (Ref. Field)	& ACS/WFC & 2$\times$350 s & F555W	& GO-12257	& L. Girardi & \\
			&			&  80 s + 300 s + 340 s  & F814W &	    		 &		       & \\\hline
    NGC 1806 (Cluster)		& ACS/WFC & 40 s + 2$\times$340 s & F555W	& GO-10595	& P. Goudfrooij & \\
			&			&  8 s + 2$\times$340 s  & F814W &	    		 &		       & \\
    NGC 1806 (Ref. Field)	& ACS/WFC & 2$\times$350 s & F555W	& GO-12257	& L. Girardi & \\
			&			&  80 s + 300 s + 340 s  & F814W &	    		 &		       & \\\hline
    NGC 1846 (Cluster)	& ACS/WFC & 40 s + 2$\times$340 s & F555W	& GO-10595	& P. Goudfrooij & \\
			&			&  8 s + 2$\times$340 s  & F814W &	    		 &		       & \\
    NGC 1846 (Ref. Field)	& UVIS/WFC3 & 2$\times$348 s & F555W	& GO-12326	& N. Keith & \\
			&			&  2$\times$400 s  & F814W &	    		 &		       & \\\hline
    NGC 1852 (Cluster)	& ACS/WFC & 330 s & F555W	& GO-9891	& G. Gilmore & \\
    			&  		    & 200 s	 & F814W &  		  & 		       & \\
    NGC 1852 (Ref. Field)	& ACS/WFC & 2$\times$500 s & F555W	& GO-12257	& L. Girardi & \\
			&			&  2$\times$350 s  & F814W &	    		 &		       & \\\hline
    NGC 2154 (Cluster)	& ACS/WFC & 300 s & F555W	& GO-9891	& G. Gilmore & \\
    			&  		    & 200 s	 & F814W &  		  & 		       & \\
    NGC 2154 (Ref. Field) & ACS/WFC & 2$\times$500 s & F555W	& GO-12257	& L. Girardi & \\
			&			&  2$\times$350 s  & F814W &	&\\\hline
    NGC 2203 (Cluster)	& UVIS/WFC3  & 120 s+2$\times$700 s & F475W & GO-12257 & L. Girardi & \\
    			&			&  30 s + 2$\times$700 s & F814W &	     &		       & \\
    NGC 2203 (Ref. Field)  & ACS/WFC & 90 s + 2$\times$500 s + 2$\times$700 s & F475W	& GO-12257	& L. Girardi & \\
			&			&  10 s + 550 s + 690 s + 2$\times$713 s  & F814W &	&\\\hline
    NGC 2213 (Cluster)	& UVIS/WFC3  & 120 s+600 s+720 s & F475W & GO-12257 & L. Girardi & \\
    			&			&  30 s + 2$\times$700 s & F814W &	     &		       & \\
    NGC 2213 (Ref. Field) & ACS/WFC & 2$\times$500 s & F475W	& GO-12257	& L. Girardi & \\
			&			&  2$\times$500 s  & F814W &	&\\\hline
			
  \end{tabular} 
  \end{center} 
   *: includes both the cluster region and the reference field.
\end{table*} 

For each cluster, we performed point-spread-function (PSF) photometry
on the flat-fielded frames ('\_flt') of the science images using the
WFC3 and ACS modules implemented in the {\sc dolphot} 2.0
package\footnote{\url{http://americano.dolphinsim.com}}. To compile
the resulting stellar catalog, we adopted a filter employing the
sharpness and `crowding' parameters calculated by {\sc
  dolphot2.0}. The sharpness illustrates the broadness of a detected
object relative to the PSF. A perfect star should have a zero
sharpness. A negative sharpness may indicate an object that is too
sharp (like a cosmic ray), while a very large positive value means
that the detected object is too broad (for example, a background
galaxy). The crowding quantifies how much brighter an object would
have been measured had nearby stars not been fitted simultaneously (in
units of magnitudes). An isolated star would have zero crowding. We
selected only objects with $-$0.3$\leq$ sharpness$\leq$0.3 and
crowding$\leq$0.5 mag in all frames used for our analysis. We only
kept objects classified as good stars, and not centrally saturated, by
{\sc dolphot2.0}. {\sc dolphot2.0} also automatically combined stellar
catalog resulting from frames with different exposure times into a
deep catalog, eventually resulting in a deep, multi-band stellar
catalog. In this paper, we only focus on evolved stars. These stars
are bright and usually have good photometric quality. We confirmed
that performing our data reduction only has a minor effect on the
stars of interest: fewer than 5\% would be removed by the data
reduction procedures adopted.

For all clusters, except NGC 1644, we have access to two parallel
observations, which represent the star cluster region and its nearby
reference field. This information is also included in Table
\ref{T1}. For NGC 1644, we simply adopted the peripheral region in the
same image, for distances to the cluster center greater than 20 pc, as
reference field. As we will show below, the best-fitting tidal radius
to the brightness profile of NGC 1644 is less than 20 pc. Our adoption
of a `field' region in the same image that contains the cluster itself
is therefore reasonable. All other clusters, except NGC 1783, have
parallel observations of reference fields located close to or well
beyond the clusters' tidal radii. NGC 1783 is so large that even a
parallel observation centered $\sim$100 pc from the cluster center
does not reach its tidal radius. However, the {\sl HST}
archive\footnote{\url{https://archive.stsci.edu/hst/search.php}} does
not contain any observations beyond the cluster's tidal radius
obtained through the same passbands. The adopted reference field for
NGC 1783 is therefore a compromise. As we will show, although we have
adopted a reference field that is actually located too close to the
cluster, field contamination in the region of interest is minor.

\begin{figure*}
\includegraphics[width=18cm]{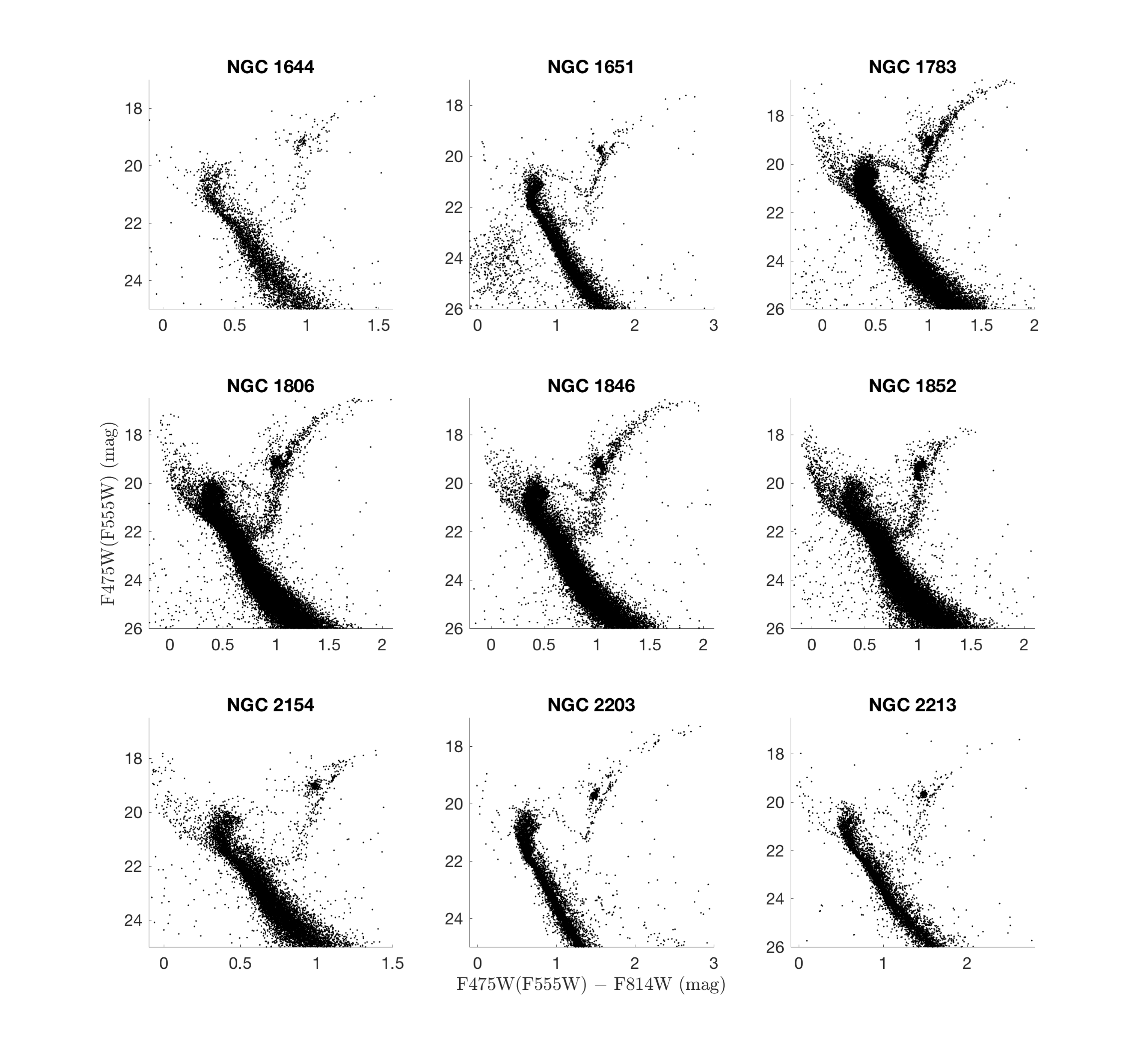}
\caption{Processed CMDs for all clusters studied in this paper.}
\label{F1}
\end{figure*}

Our cluster CMDs are presented in Fig. \ref{F1}.

Before we search for eBSS candidates in our sample clusters, we first
calculated their brightness profiles, which will allow us to derive
the clusters' structural parameters (e.g. core, half-light, and tidal
radii). To obtain reliable structural parameters, we first calculated
the clusters' center coordinates. After transferring the CCD
coordinates $(X,Y)$ to equatorial coordinates ($\alpha_{\rm J2000}$,
$\delta_{\rm J2000}$) for all detected stars, we calculated the
stellar number density contours for each cluster. We then defined the
position where the stellar number density reaches its maximum value as
the cluster center. For each cluster, we selected all stars brighter
than a given magnitude in the F814W filter as a subsample. This
magnitude limit is about two or three magnitudes brighter than the
detection limit. We then used this subsample to study the clusters'
brightness profiles. We only selected these bright sample stars
because (1) these stars will have a high completeness level and (2) in
a stellar system, massive stars contribute most of the flux.

\begin{figure*}
\includegraphics[width=18cm]{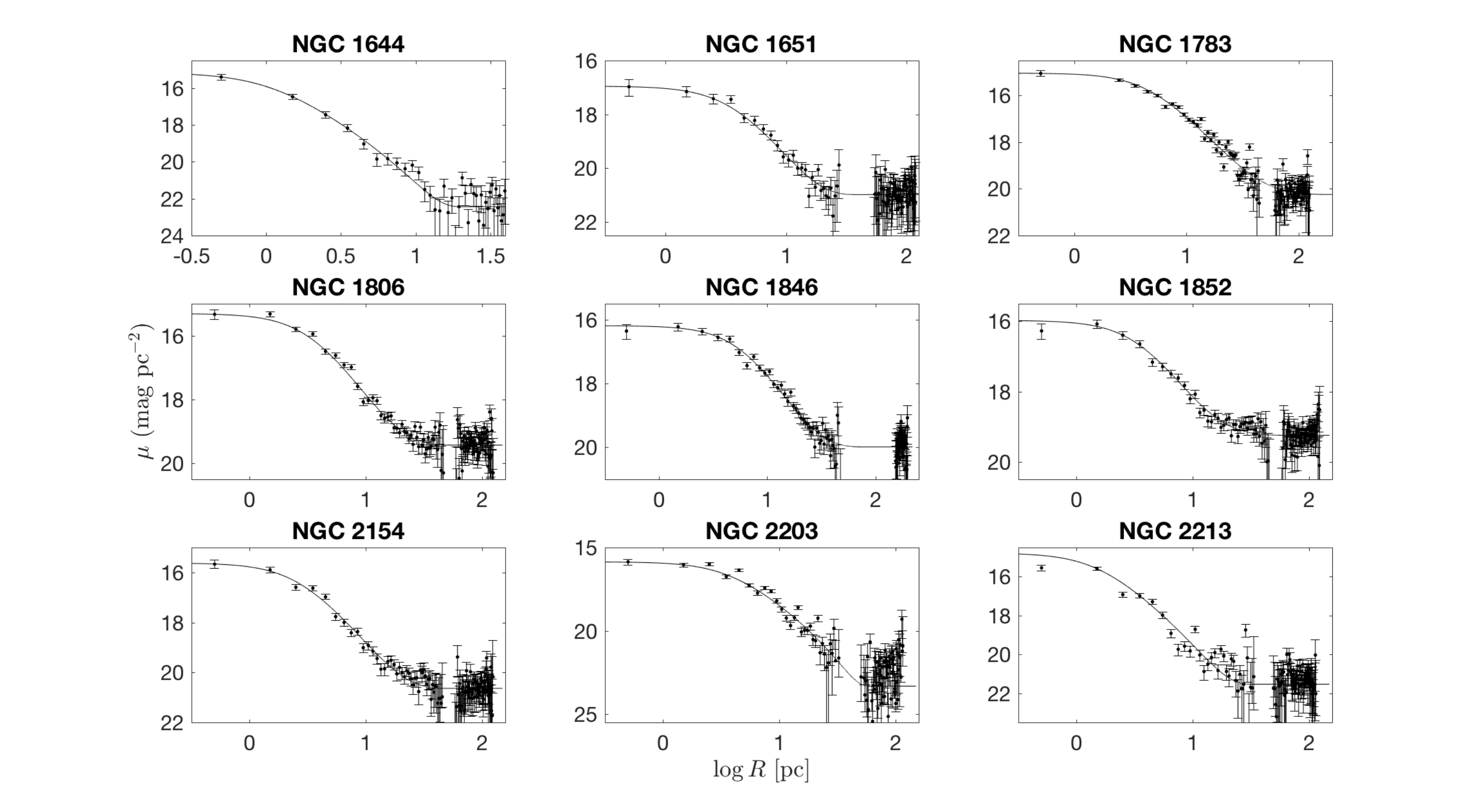}
\caption{Brightness profiles (in the F814W filter) of our sample
  clusters. The solid curves represent the best-fitting King models.}
\label{F2}
\end{figure*}

We used the cluster center to define different annular rings. This
approach was applied to the observations of both the cluster image and
the reference field. We adopted radial intervals between each pair of
successive annular rings of 1 pc. For each ring, we calculated the
total flux of stars (in the F814W filter), $f(r)=\sum{10^{({\rm
      F814W}-(m-M)_0/(-2.5)}}$. The flux density is
$\rho(r)=f(r)/A(r)$, corresponding to a surface brightness of
$\mu(r)=-2.5\log{\rho(r)}+(m-M)_0$. In principle, the total flux in
each ring should contain the contributions of both cluster as well as
field stars. This brightness profile can be described by a King model
combined with a constant (representing the field brightness),
\citep{King62a}.
\begin{equation}
\mu(r)=k\left[\frac{1}{\sqrt{1+(r/r_{\rm
        c})^2}}-\frac{1}{\sqrt{1+(r_{\rm t}/r_{\rm c})^2}}\right]+{b}.
\end{equation}
Here, $r_{\rm c}$ and $r_{\rm t}$ are the core and tidal radii,
respectively; $b$ is a constant which represents the background level,
and $k$ is a normalization coefficient. The derived brightness
profiles for our clusters, as well as their best-fitting King models,
are presented in Fig. \ref{F2}. We have assumed that the uncertainties
in the brightness are Poisson-like.

We calculated the clusters' half-light radii, $r_{\rm hl}$, based on
their best-fitting King models. The derived center coordinates for our
clusters, as well as their best-fitting structural parameters ($r_{\rm
  c}$, $r_{\rm hl}$, and $r_{\rm t}$), are presented in Table
\ref{T2}.

\begin{table*}
  \begin{center}
\caption{Derived structural parameters. Second to sixth columns:
  center right ascension and declination, core, half-light, and tidal
  radii, and median distance to the cluster center for the reference
  field stars.}\label{T2}
  \begin{tabular}{c c c c c c rl}\hline
    Cluster      &  $\alpha_{\rm J2000}^*$ (deg) & $\delta_{\rm J2000}^*$ (deg) & $r_{\rm c}$ (pc) & $r_{\rm hl}^{\S}$ (pc) & $r_{\rm t}$ (pc) & $\widetilde{r_{\rm f}}^{\dag}$ (pc) \vspace{1mm}\\\hline
    NGC 1644 	& $04^{\rm h}37^{\rm m}39.84^{\rm s}\pm0.48^{\rm s}$ & $-66^{\circ}11'56.40''\pm3.60''$ & 1.04$\pm$0.01 &2.34$^{+0.16}_{-0.74}$ &19.57$\pm$0.52 & 28.10\\
    NGC 1651 	& $04^{\rm h}37^{\rm m}31.80^{\rm s}\pm0.48^{\rm s}$ & $-70^{\circ}35'07.08''\pm3.60''$  & 3.59$\pm$0.11 &5.86$^{+0.64}_{-0.36}$ &36.64$\pm$4.91 & 88.32\\
   NGC 1783	& $04^{\rm h}59^{\rm m}08.88^{\rm s}\pm0.60^{\rm s}$ & $-65^{\circ}59'13.20''\pm3.60''$  & 4.68$\pm$0.13 &13.82$^{+0.68}_{-0.32}$ &181.20$\pm$115.17 & 95.95\\
   NGC 1806	& $05^{\rm h}02^{\rm m}12.12^{\rm s}\pm0.60^{\rm s}$ & $-67^{\circ}59'07.80''\pm3.60''$  & 3.39$\pm$0.09 &6.98$^{+0.52}_{-0.48}$ &52.60$\pm$10.35 & 94.22\\
   NGC 1846	& $05^{\rm h}07^{\rm m}34.68^{\rm s}\pm0.48^{\rm s}$ & $-67^{\circ}27'32.40''\pm3.60''$  & 5.81$\pm$0.11 &10.72$^{+0.78}_{-0.22}$ &73.84$\pm$8.08 & 175.59\\
   NGC 1852	& $05^{\rm h}09^{\rm m}23.76^{\rm s}\pm0.36^{\rm s}$ & $-67^{\circ}46'48.00''\pm3.60''$  & 3.64$\pm$0.07 &7.57$^{+0.93}_{-0.07}$ &61.11$\pm$8.88 & 93.31\\
   NGC 2154	& $05^{\rm h}57^{\rm m}38.16^{\rm s}\pm0.36^{\rm s}$ & $-67^{\circ}15'46.80''\pm3.60''$  & 2.61$\pm$0.08 &5.72$^{+0.78}_{-0.22}$ &48.00$\pm$12.31 & 92.61\\
   NGC 2203	& $06^{\rm h}10^{\rm m}42.24^{\rm s}\pm0.36^{\rm s}$ & $-71^{\circ}31'44.76''\pm3.60''$  & 3.62$\pm$0.08 &7.51$^{+0.99}_{-0.01}$ &59.33$\pm$10.44 & 84.58\\
   NGC 2213	& $06^{\rm h}10^{\rm m}42.24^{\rm s}\pm0.48^{\rm s}$ & $-71^{\circ}31'44.76''\pm3.60''$  & 1.48$\pm$0.04 &3.53$^{+0.97}_{-0.03}$ &31.31$\pm$2.18 & 84.05\\
   \hline
  \end{tabular} 
  \end{center} 
  $*$: The uncertainty is equal to the size of the spatial bins
  adopted to calculate the number density contours.\\ 
  $\S$: The uncertainty is set equal to the radial interval (1 pc
  between two radial bins).\\
  $\dag$: Median distance to the cluster center for all stars in the
  reference field.
\end{table*} 

Here we see that, except for NGC 1644, all clusters have tidal radii
of at least $\sim$30 pc. For comparison, the field of view (FOV) of
ACS/WFC is $202''\times202''$; at the distance of the LMC, this is
equal to a square with an area of $\sim48.5\times48.5$ pc$^2$. The
clusters NGC 1651, NGC 1718, NGC 2203, and NGC 2213 were observed with
the UVIS/WFC3. Its FOV is even smaller, $162''\times162''$
($\sim38.9\times38.9$ pc$^2$). The tidal radius derived for NGC 1644
is only $\sim$20 pc, so that adopting the edge of the image as
reference field is therefore reasonable. For the other clusters, we
use the parallel observations as their reference fields. In Table
\ref{T2}, we also list the median distances of stars in the reference
fields to the cluster centers. Because of the large sizes of our
clusters, in Fig. \ref{F1}, most CMDs contain stars from across the
full cluster images. For NGC 1644, however, the CMD of the cluster
region contains only stars within 20 pc from the cluster center.

\section{Main Results}\label{S3}

\begin{figure*}
\includegraphics[width=18cm]{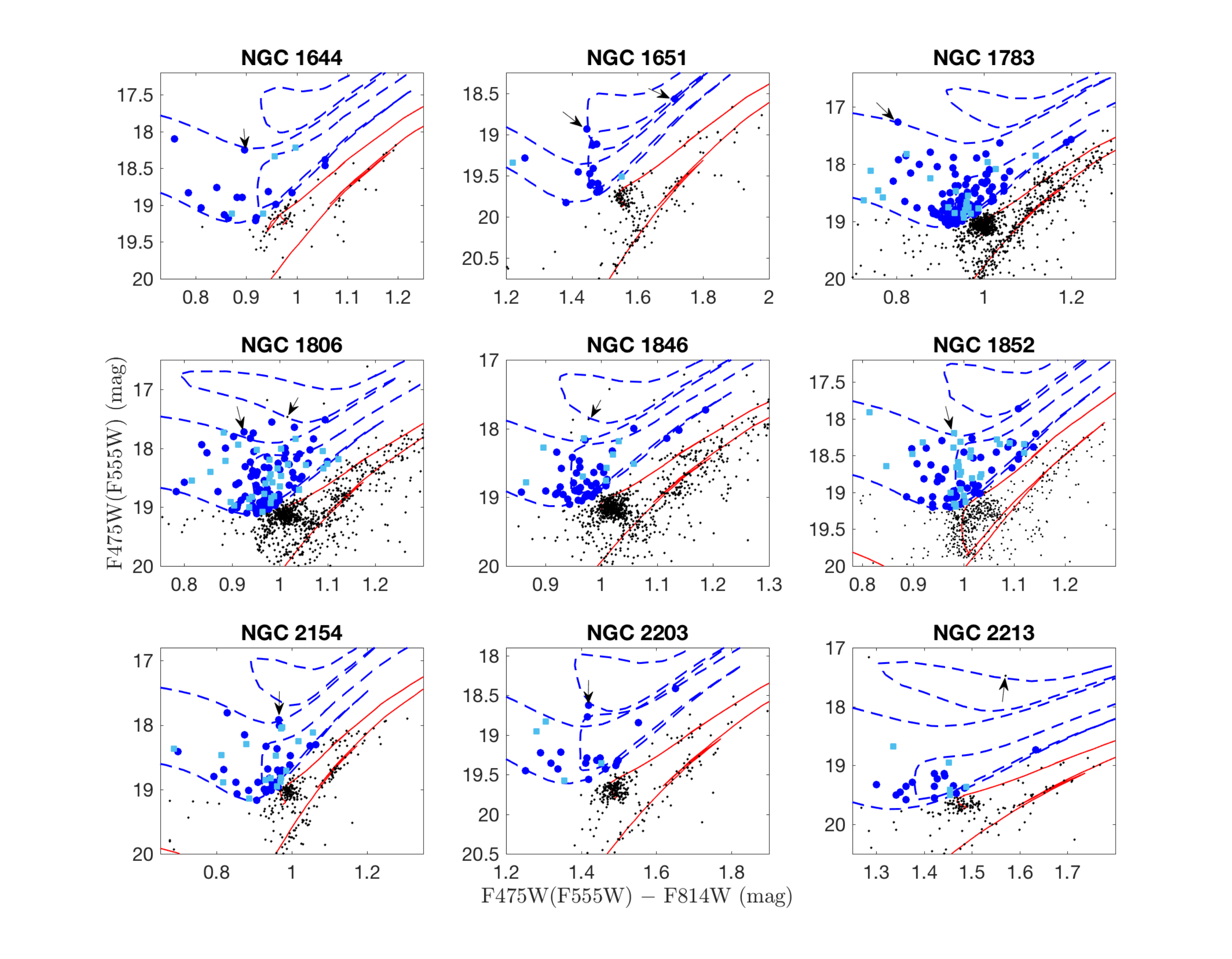}
\caption{CMDs of our clusters, zoomed into the RC regions. All eBSS
  candidates observed in the cluster regions are indicated by dark
  blue circles, while field-star contaminators are shown as light blue
  squares. The red solid lines are the best-fitting isochrones for the
  bulk stellar populations. Blue solid lines are the adopted young
  isochrones which roughly define the magnitude boundaries for
  selecting the eBSSs. Arrows indicate stars that were used to
    determine the top boundaries (through visual inspection).}
\label{F3}
\end{figure*}

In Fig. \ref{F3} we present the CMDs of our nine clusters, zoom into
the regions centered on their RCs. All clusters contain a sample of
stars located to the bright and blue side of the RC and the AGB. These
stars are likely evolved stars, although they are more massive than
normal giant stars. To select these stars we first use the PARSEC
isochrones to fit our observations based on visual inspection
\citep{Bres12a}. For each cluster, we use an isochrone with a older
age to fit the bulk stellar population, while we use two younger
isochrones to roughly describe the bottom and top boundaries of the
sample of bright evolved stars. The bottom boundary is the
  isochrone that crosses the bluer and brighter side of the RC in the
  cluster of interest, chosen to avoid selecting too many stars that
  are affected by photometric scatter. These bottom boundaries should
  go across the region where the stellar number density is
  significantly lower than that in the RC. For the top boundary, we
  tried different isochrones pertaining to a range of younger ages to
  fit these bright evolved stars. We chose the isochrone that best
  covered most of the brightest stars in our sample clusters as the
  top boundary. Stars used to determine the top boundaries are
  indicated by arrows in Fig. \ref{F3}. The adopted parameters are
presented in Table \ref{T3}. Our best-fitting parameters are close to
or consistent with those of \cite{Milo09a,Li14a,Nied16a}.

For each cluster, we first adopted stars that are redder than the
middle point of the subgiant branch (SGB) as our sample stars, based
on the adopted two young isochrones. Stars that are covered by the top
and bottom boundaries are identified as bright evolved stars and
  will be used for analysis. In Fig. \ref{F3}, the selected bright
evolved stars are highlighted by dark blue circles. For each cluster,
the same selection method was also applied to the CMDs of their
reference fields. The field contaminators are highlighted by light
blue squares.

However, except for NGC 1644, we cannot directly compare the
  observed bright evolved stars with their counterparts in the
  reference fields. This is because the observed fields usually have
  different exposure times than the cluster regions. Completeness
  levels in the same part of the CMDs between the reference field and
  the cluster are therefore usually different. For most of our
  clusters, the reference field observations do not include images
  with short exposure times. Bright evolved stars observed in the
  cluster region may be saturated in the reference field. In such
  cases, we may not be able to quantify how many field stars could
  contaminate this very bright CMD region. Therefore, we removed some
  bright evolved stars from the cluster sample so as to conduct a fair
  comparison. We generated a sample of artificial stars with same
  magnitudes as our selected bright evolved stars. We then added all
  of these artificial candidates to the reference field and recovered
  them using {\sc dolphot2.0}. Finally, we removed those artificial
  stars that would be saturated in the observed reference fields. For
  this reason, there are some very bright evolved stars in some panels
  of Fig. \ref{F3} that were not selected for further analysis (but
  they were used to determine the youngest isochrones used for the
  fitting boundaries).

\begin{table*}
  \begin{center}
\caption{\bf Best-fitting parameters of the adopted isochrones}\label{T3}
  \begin{tabular}{c c c c c c c c c c c c}\hline
    Cluster      &  log $t_{\rm 1}$ & log $t_{\rm 2}$ & log $t_{\rm 3}$ & log $t_{\rm 2M_1}$ & $M_1$ & $M_2$& $M_3$& $M_3/(2M_1)$ & $Z^*$& $A_{\rm V}$& $(m-M)_{0}$\\
    	&	[yr]	&	[yr]	&	[yr]	& [yr] & $(M_{\odot})$	&$(M_{\odot})$&$(M_{\odot})$& & & (mag) & (mag) \\
	& (1) & (2) & (3) & (4) & (5) & (6) & (7) & (8) & (9) & (10) & (11) \\\hline
    NGC 1644 	& 9.20 & 8.85 & 8.57 & 8.40 & 1.70 & 2.27 & 2.90 & 85.29\% & 0.008 &0.03$\pm0.02$ & 18.50 \\
    NGC 1651 	& 9.26 & 8.83 & 8.70 & 8.45 & 1.59 & 2.27 & 2.54 & 79.87\% & 0.006 &0.31$\pm0.02$ & 18.50 \\
    NGC 1783 	& 9.23 & 8.80 & 8.40 & 8.43 & 1.62 & 2.32 & 3.33 & 102.78\% & 0.006 &0.06$\pm0.02$ & 18.50 \\
    NGC 1806 	& 9.25 & 8.80 & 8.47 & 8.45 & 1.55 & 2.26 & 3.05 & 98.39\% & 0.004 &0.24$\pm0.04$ & 18.50 \\
    NGC 1846 	& 9.22 & 8.80 & 8.54 & 8.42 & 1.64 & 2.32 & 2.93 & 89.33\% & 0.006 &0.15$\pm0.03$ & 18.50 \\
    NGC 1852 	& 9.10 & 8.80 & 8.53 & 8.31 & 1.80 & 2.32 & 2.99 & 83.06\% & 0.006 &0.24$\pm0.03$ & 18.50 \\
    NGC 2154 	& 9.21 & 8.83 & 8.50 & 8.41 & 1.65 & 2.27 & 3.04 & 92.12\% & 0.006 &0.09$\pm0.02$ & 18.50 \\
    NGC 2203 	& 9.22 & 8.80 & 8.60 & 8.42 & 1.64 & 2.32 & 2.78 & 84.76\% & 0.006 &0.22$\pm0.02$ & 18.50 \\
    NGC 2213 	& 9.25 & 8.85 & 8.45 & 8.44 & 1.60 & 2.23 & 3.18 & 99.38\% & 0.006 &0.19$\pm0.03$ & 18.50 \\    \hline
  \end{tabular} 
  \end{center} 
  (1) Age of the bulk population stars (in logarithmic units). (2, 3)
  Upper, lower limits to the ages of the young evolved stars. (4) Age
  of the population with twice the mass of the bulk stellar
  population. (5) Mass of bottom RGB star of the bulk
  population. (6, 7) Lower, upper limits to the masses of the bright
  evolved stars. (8) Mass ratio of the upper mass limits to twice the
  mass for the bottom RGB star. (9) Metallicity. (10) Reddening and
  internal spread (differential reddening). (11) Distance
  modulus.\\
  $*$: $Z_{\odot}=0.0152$
\end{table*} 

However, simply using isochrones to define bright evolved stars is not
reliable. Because real observations always contain scatter and field
contamination, we must examine if these bright evolved stars are
simply caused by effects of photometric artifacts or by field
contamination.

\subsection{Photometric Scatter, Distance Spread, and Differential Reddening}

In real observations, numerous photometric effects may introduce
scatter: (a) Photometric uncertainties will broaden evolutionary
patterns (i.e., RGB, RC, and AGB) for normal evolved stars. Very large
photometric uncertainties may scatter normal stars into the adopted
region for bright evolved stars, erroneously making us treat them as
massive stars. (b) Stars may occasionally blend with other stars or
with their diffraction patterns, with extended sources (such as
background galaxies), cosmic rays, and/or bad pixels. These crowding
effects will change the positions of stars in their CMD.

The distances to stars in a star cluster are not exactly identical. In a
star cluster, stars at slightly different distances will produce a
vertical structure across the RC in their CMD \citep{Gira16a}. For
example, if we assume that most stars in NGC 1783 are located at its
tidal radius ($\sim$180 pc; see Table \ref{T2}), at the distance of
the LMC (50 kpc) this will produce a magnitude spread of up to
$\sim$0.016 mag in each
passband\footnote{$\Delta{M}=5\log{50180}-5\log{49820}$}. 

A cluster region may contain many dusty clumps, which will produce an
inhomogeneous reddening distribution across the whole cluster
region. The differential reddening may scatter normal stars into the
regions of bright evolved stars as well. The typical reddening spread
for many old GGCs is $\Delta{E(B-V)}\sim-0.04$ to 0.04 mag
\citep{Milo12a}. Differential reddening for Magellanic Cloud clusters
is not well-studied. \cite{Mart17a} investigated the differential
reddening level in the region of NGC 419, a Small Magellanic Cloud
(SMC) cluster with a similar age as our sample clusters. They
concluded that the reddening in the F336W filter ($A_{\rm F336W}$)
does not exceed 0.02 mag, which roughly correponds to a maximum
reddening spread of $\Delta{E(B-V)}=0.004$ mag. However,
\cite{Zhang18a} studied the extinction map for the cluster region of
NGC 1783 using the same method as \cite{Milo12a}, concluding that the
maximum extinction variation across the cluster region ranges from
$\Delta{E(B-V)}\sim-0.03$ to 0.06 mag.

Effects (a) and (b) can be studied through artificial star (AS)
tests. The principle underlying AS tests is to generate a sample of
fake stars using the same point-spread-function (PSF) as for the real
stars\footnote{Because space observations are not affected by
  atmospheric seeing, {\sc HST} PSFs for individual stars only depend
  on their (known) position on the CCD.}, and then input these fake
stars into the raw image and recover them using the same method as
applied to the real stars.

The effects of differential reddening can be explored by
  examining the shape of the RC. A dusty cluster will exhibit an
  elongated RC (as well as other elongated features) along the
  reddening direction. Comparing mock observations produced based on
  AS tests, which only included broadening by photometric artifacts,
  to real observations helped us quantify the possible level of any
  differential reddening.

To study the combination of these effects, for each
cluster we simulated a population of normal evolved stars. We
generated these simulated stellar populations following four steps:

\begin{enumerate}
\item Generate a sample of ASs with a Kroupa-like mass function
  \citep{Krou01a}. All ASs are initially located on the best-fitting
  isochrone. In order to derive a statistically robust result for
  estimating the effect of photometric scatter, the total number of
  ASs is more than 20 times that of the observations.
\item Assume a Gaussian distribution of distances for all stars
  between $d-r_{\rm t}$ and $d+r_{\rm t}$, where $d$ is the central
  distance to the cluster and $r_{\rm t}$ is its best-fitting tidal
  radius. Then correct individual stars for the differences in
  distance modulus.
\item Input all ASs into the raw image and recover them using the same
  method as applied to the real observations. We repeated this process
  many times to avoid blending between two ASs; each time we only
  input 100 ASs into the raw image.
\item For each star, we randomly assigned an extinction variation,
  derived from a Gaussian distribution of $\Delta{E(B-V)}$.We then
  corrected their magnitudes in each passband using the \cite{Card89a}
  extinction curve with $R_{V} =3.1$. $\Delta{E(B-V)}$ was
    determined by comparing the resulting mock CMDs to real
    observations.
\end{enumerate}

\begin{figure}
\includegraphics[width=9cm]{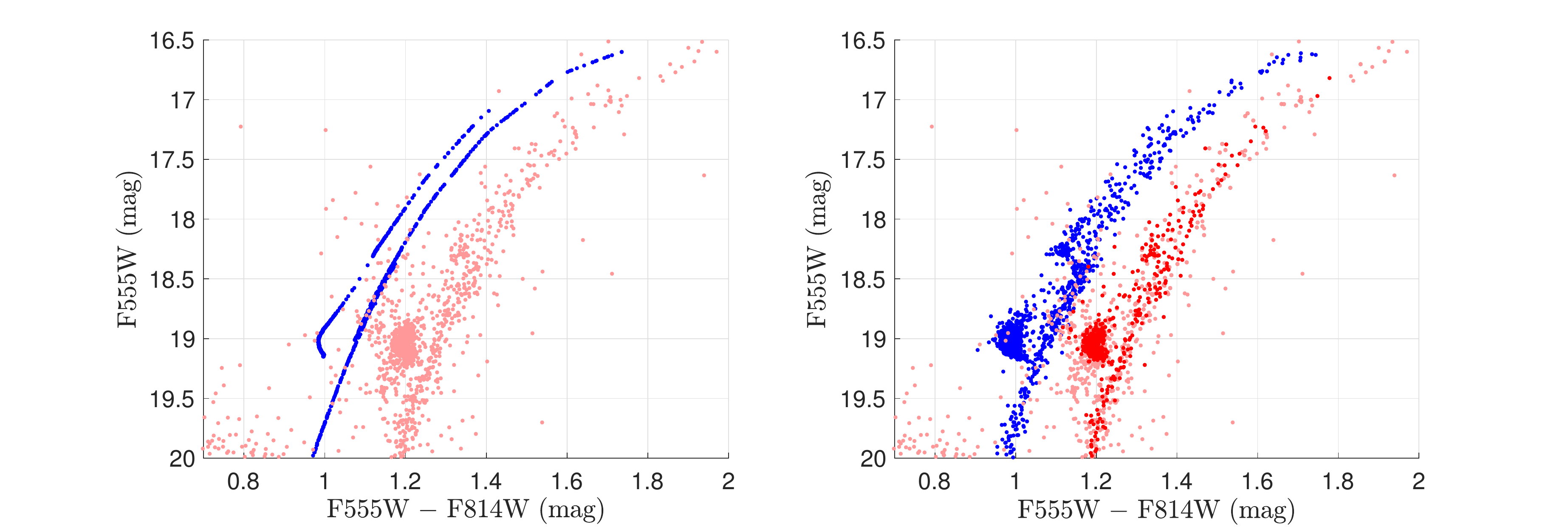}
\caption{Comparison of a simulated, evolved stellar population
  with the real observations of NGC 1783. Blue and red dots represent
  simulated and observed stars, respectively. The observed CMD has
  been shifted by 0.2 mag in color. The left-hand panel shows input
  ASs not affected by any scatter; the right-hand panel shows the
  output artificial stars, which are affected by photmetric artifacts,
  a distance spread, and differential reddening. Dark red dots
  represent successfully modeled stars.}
\label{F4}
\end{figure}

In Fig. \ref{F4} we show a comparison of the simulated, evolved
  stellar population with the real observations of NGC 1783. The main
  features of the RGB, AGB, and RC of NGC 1783 are well reproduced by
  our simulation. The derived differential reddening levels for our
  sample clusters are indicated by the reddening uncertainties in
  Table \ref{T3}.

\begin{figure*}
\includegraphics[width=18cm]{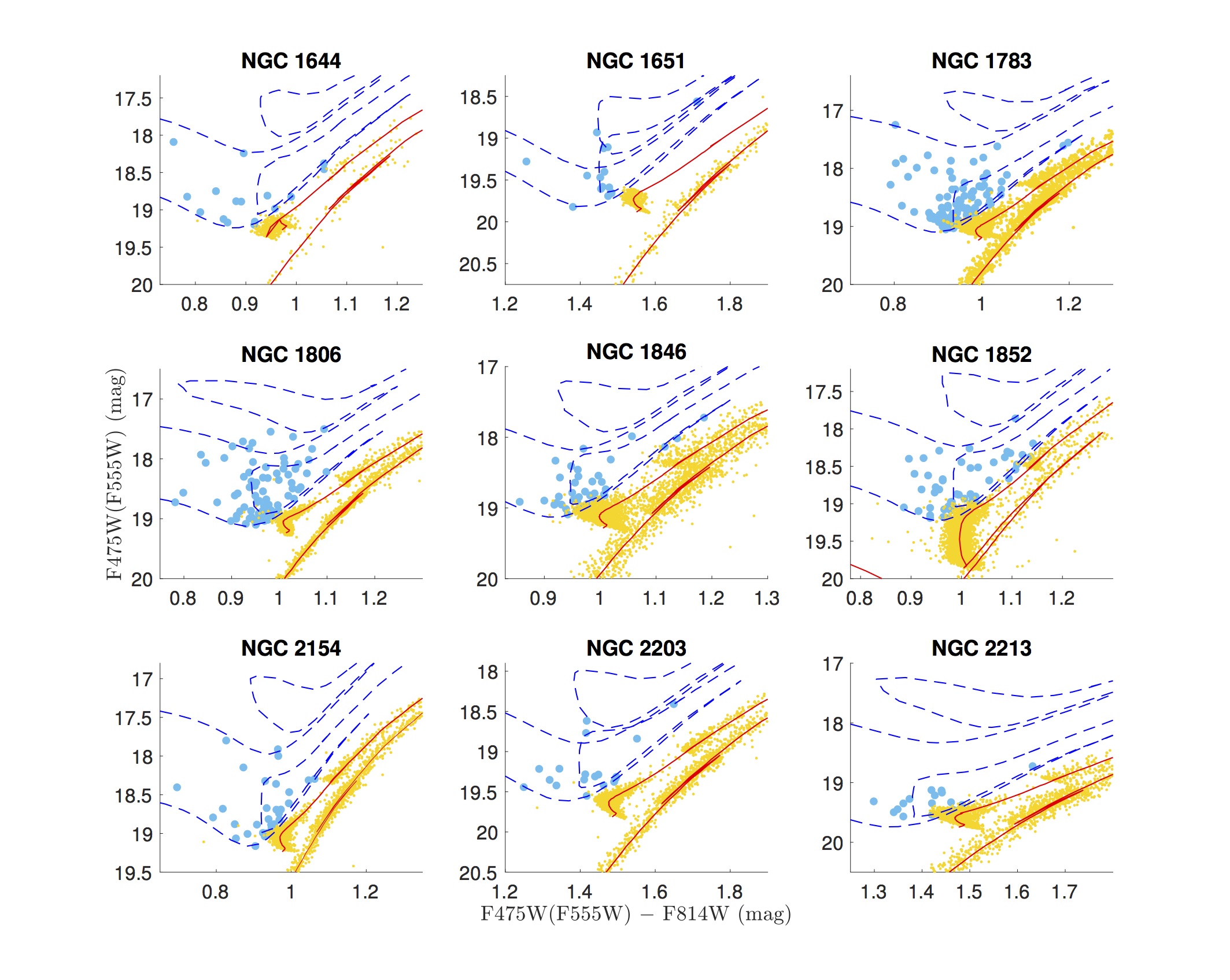}
\caption{CMDs of the bright evolved stars (blue circles) as well as
  the simulated normal evolved stars (orange dots). The blue and red
  solid lines are isochrones adopted for fitting the normal population
  stars and to define the boundaries of the bright evolved-star
  regions, respectively.}
\label{F5}
\end{figure*}

We then directly compared the observed CMDs with our simulations. Here
we simply wanted to examine if all these effects will scatter many
normal evolved stars into the region occupied by the bright evolved
stars. In Fig. \ref{F5} we show the CMDs of both the selected bright
evolved stars and the simulated normal stars for our clusters.

As shown in Fig. \ref{F5}, a small fraction of stars indeed scatter
into the region of the bright evolved stars. We found that most
  of these scattered stars are RC stars. To further quantify the
  combined effects of photometric artifacts and any internal distance
  and reddening spreads, we calculated the number ratio of ASs that
  were scattered into the eBSS region to all artificial evolved stars,
  which we call the scatter probability. Next, we estimated the number
  of normal evolved stars whose distances and reddening properties can
  be adequately modeled by our artificial stellar populations. To do
  this, we evaluated the distance between all observed stars and their
  nearest AS in the CMD. For each observed star, if we could find an
  AS within the $3\sigma$ photometric uncertainty, we treated it as a
  successfully modeled star. The number difference of successfully
  modeled stars between the cluster region and the reference field
  defines the expected number of normal evolved stars in the star
  cluster. These numbers multiplied by the scatter probabilities
  determine the expected numbers of scattered stars in the real
  observations, which are listed in Table \ref{T4}. Our results show
  that the expected number of stars that are scattered into the eBSS
  region varies from less than one (NGC 1644, NGC 1651, and NGC 2154)
  to 15 (NGC 1852). However, these numbers are all significantly
  smaller than the observed numbers of bright evolved stars in our
  clusters. In addition, the color--magnitude distributions of these
contaminating stars are centered close to the RC. In contrast, the
observed bright evolved stars are more dispersed, with many of them
found close to the upper boundary. We therefore conclude that the
populations of observed bright evolved stars cannot be fully explained
by the scatter of measurement.

\subsection{Field Contamination}

Because the LMC is a star-forming galaxy, its field population
contains many young stars that may contaminate our observed
clusters. However, because of the large distance to the LMC and the
crowded environments of its star clusters, direct examination if an
individual star observed in any of our clusters is a member star
(based on proper motion or radial velocity analysis) is not
possible. A frequently used method is to compare the cluster's CMD to
the CMD of its nearby field. If the number of observed bright evolved
stars in the cluster region is significantly larger than that in the
reference field (per unit area), the normalized number difference
should be statistically equal to the number of genuine cluster
members. In Table \ref{T4} we present the observed numbers of bright
evolved stars in the cluster regions ($N_{\rm cl}$) and the reference
fields ($N_{\rm f}$), as well as the expected number of field
contaminators in the cluster regions ($N_{\rm con}$) after correcting
for the difference in observational area. We list the area ratio of
the reference field and the cluster region as well.

\begin{table*}
  \begin{center}
\caption{Expected numbers of scattered stars in the real
  observations. Following the cluster name (1), the first data block
  lists the observed number of bright evolved stars in the cluster (2)
  and in the reference field (3), the expected number of contaminating
  field stars (4), and the area ratio of the reference field to the
  cluster region (5). The second data block contains the number of
  modeled giant stars (6), the expected number of scattered stars (7),
  and the scatter probability (8). }\label{T4}
  \begin{tabular}{c | c c c c | c c r}\hline
    Cluster      &  $N_{\rm cl}$ & $N_{\rm f}$ & $N_{\rm cont}$ & $A_{\rm f}/A_{\rm c}$ & $N_{\rm giant}$ & $N_{\rm scat}$ & $P_{\rm sca}$ \\
   (1) & (2) & (3) & (4) & (5) & (6) & (7) & (8)\\\hline
    NGC 1644 & 17 & 4 & $\sim$5 & 0.79 & $\sim$41& $<$1 & 0.56\%\\
    NGC 1651 & 15 & 4 & 2--3 & 1.55 & $\sim$113 & $<$1 & 0.40\% \\
    NGC 1783 & 100 & 23 & 23 & 1 &919 & $\sim$6 & 0.70\%\\
    NGC 1806 & 86 & 34 & 34 & 1 &565 & $\sim$2 &0.40\% \\
    NGC 1846 & 43 & 9 & 13--14 & 0.65 &784--785 &$\sim$12 & 1.50\%\\
    NGC 1852 & 56 & 30 & 30 & 1 &394 & $\sim$13 & 3.35\% \\
    NGC 2154 & 33 & 16 & 16 & 1 & 205 & 1--2 & 0.73\% \\
    NGC 2203 & 18 & 8 & 5--6 & 1.55 &218--219 & $<$1 & 0.07\%\\
    NGC 2213 & 16 & 8 & 5--6 & 1.55 &$\sim$92 & 4--5 & 4.93\% \\\hline
  \end{tabular} 
  \end{center} 
\end{table*} 

We find that for all our clusters, the observed numbers of bright
evolved stars in the cluster regions are at least twice those in the
reference fields (except for NGC 1852). These differences are
statistically significant, which proves that for our clusters, at
least $\sim50$\% of the observed eBSS candidates are likely genuine
cluster members.

To further verify the cluster membership probabilities of the observed
bright evolved stars, a promising approach is to check their spatial
distributions. Because field stars are not gravitationally bound to
the cluster, their spatial distribution should be roughly
homogeneous. Genuine populations of cluster member stars should
exhibit a clear central concentration.

\begin{figure}
\includegraphics[width=9cm]{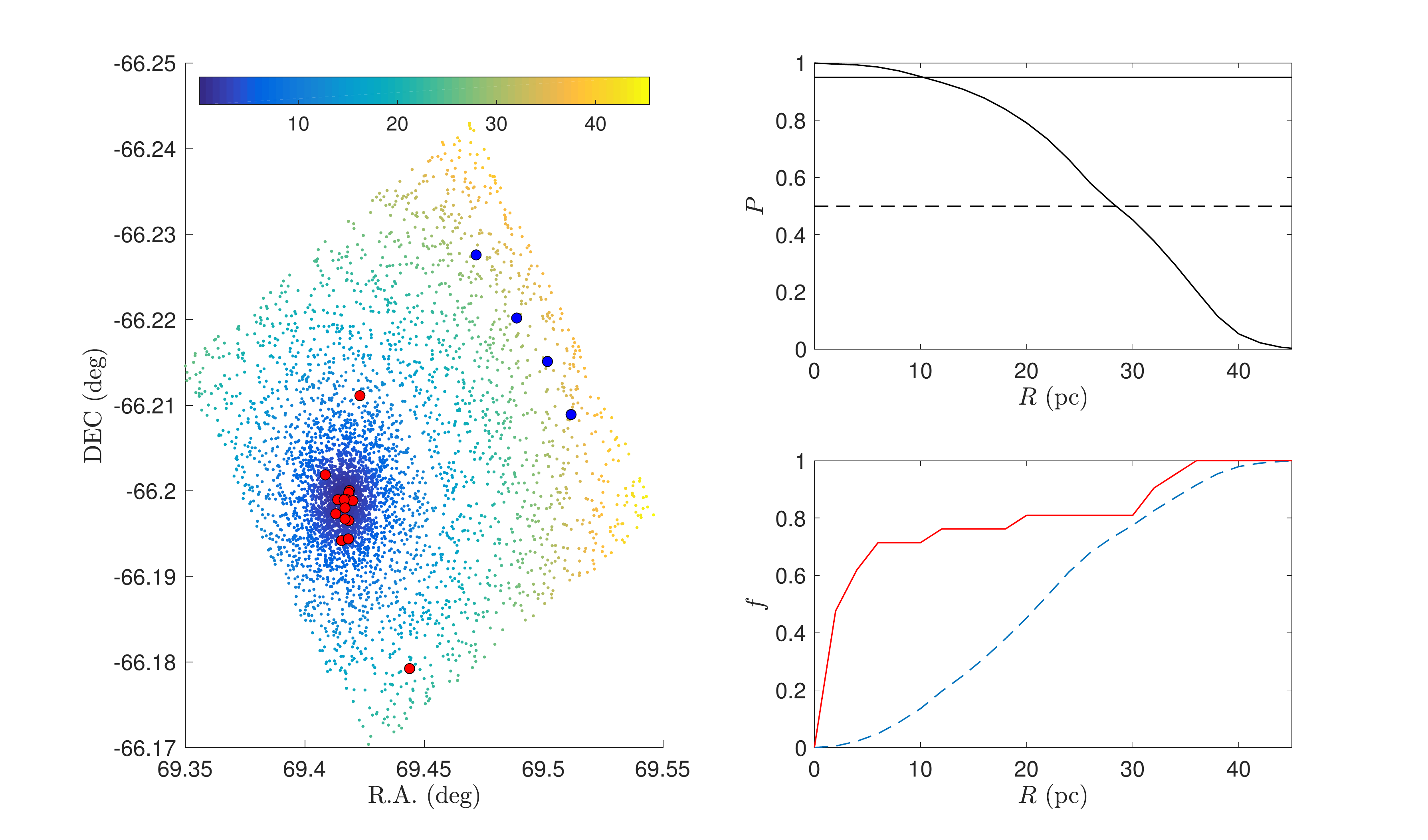}
\caption{(left) Spatial distributions of all stars in NGC 1644, the
  selected bright evolved stars (red circles), and their corresponding
  field contaminators (blue circles). The colors of the background
  points indicate their 2D distance to the cluster center. (top right)
  Probability that stars inside different radii are produced by
  genuine cluster stars. The solid and dashed horizontal lines
  indicate significance levels of 95\% and 50\%, respectively. (bottom
  right) Cumulative profiles of the observed bright evolved stars (red
  solid lines) and simulated field stars (blue dashed lines).}
\label{F6}
\end{figure}

\begin{figure}
\includegraphics[width=9cm]{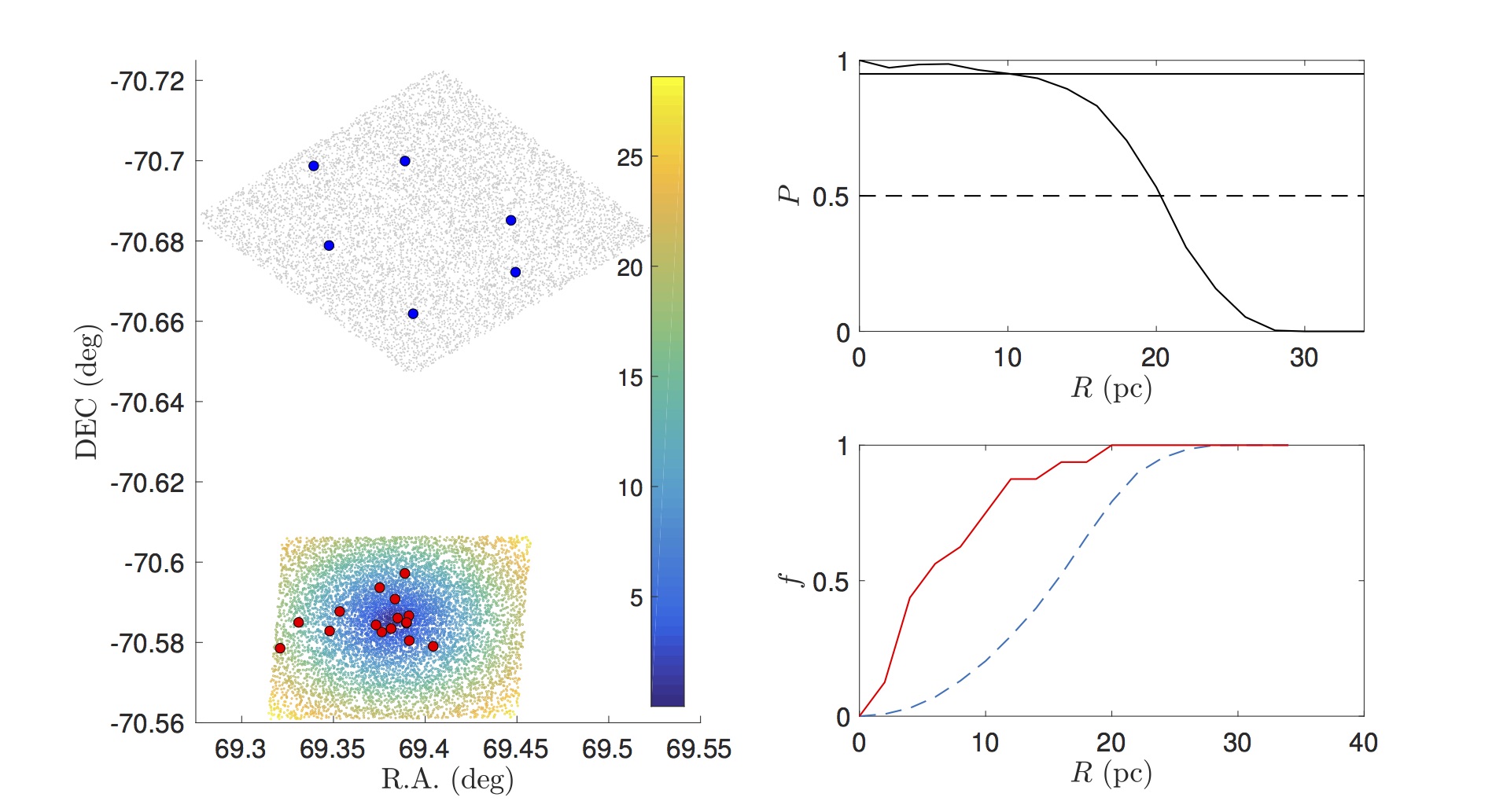}
\caption{As Fig. \ref{F6} but for NGC 1651.}
\label{F7}
\end{figure}

\begin{figure}
\includegraphics[width=9cm]{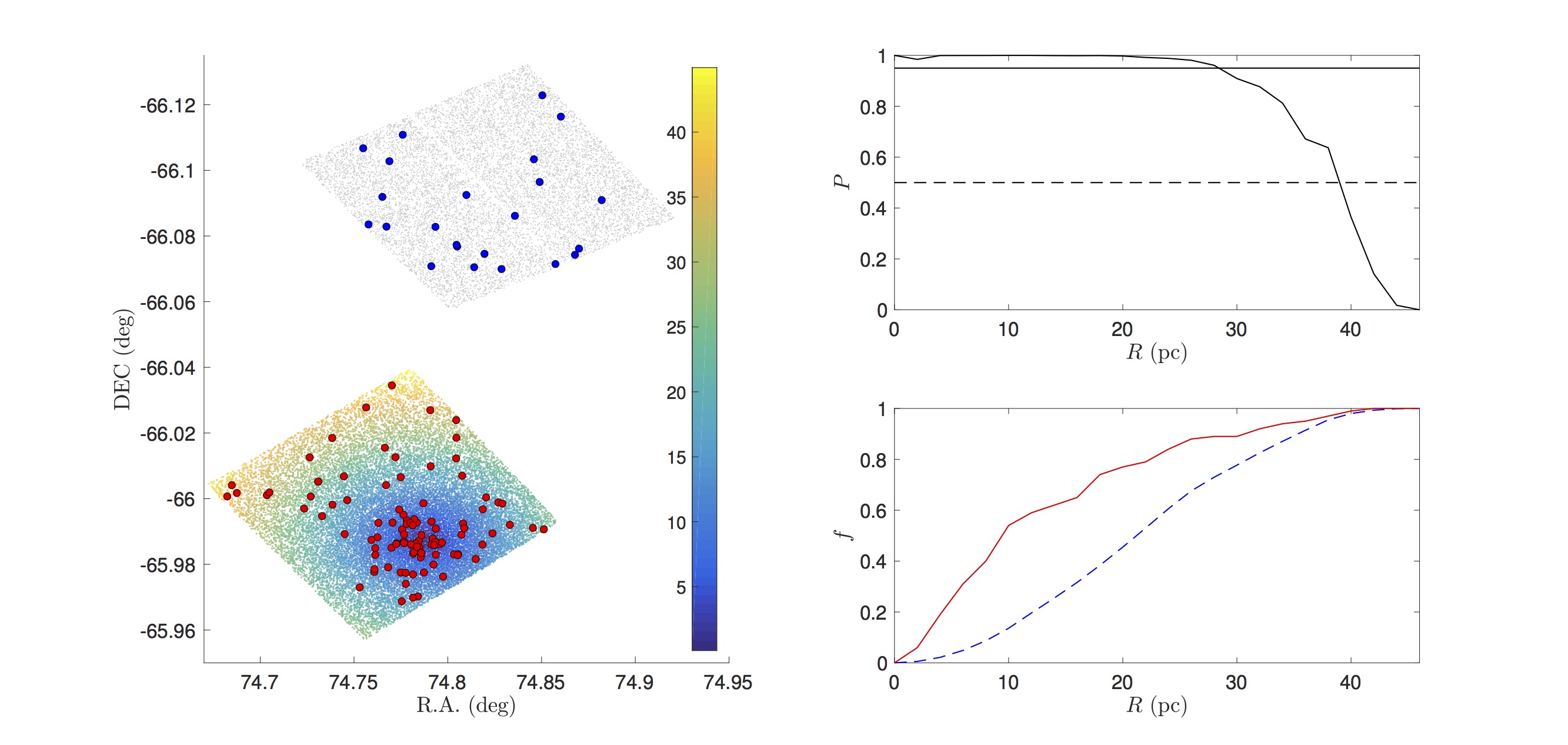}
\caption{As Fig. \ref{F6} but for NGC 1783.}
\label{F8}
\end{figure}

\begin{figure}
\includegraphics[width=9cm]{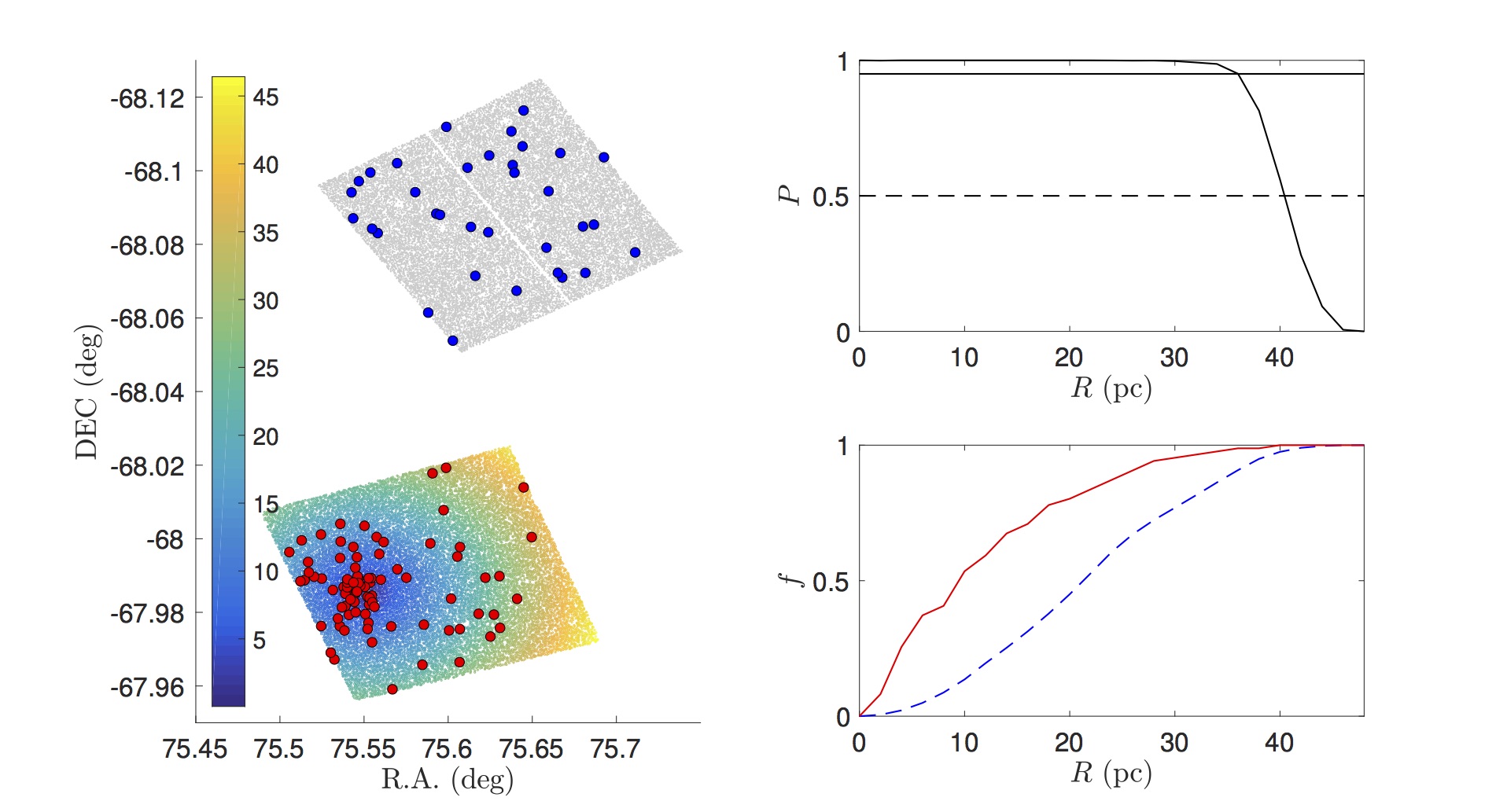}
\caption{As Fig. \ref{F6} but for  NGC 1806.}
\label{F9}
\end{figure}

\begin{figure}
\includegraphics[width=9cm]{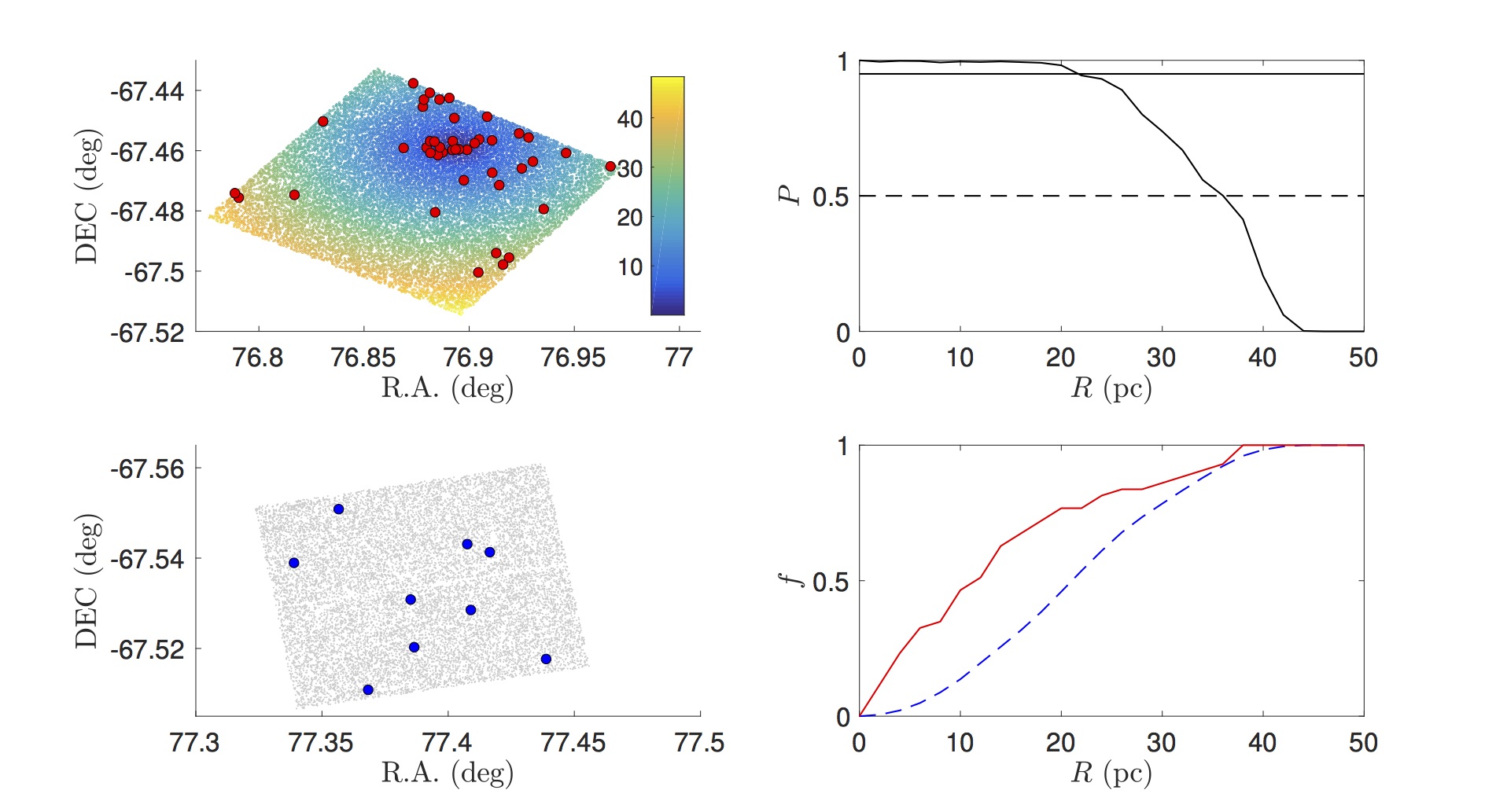}
\caption{As Fig. \ref{F6} but for NGC 1846.}
\label{F10}
\end{figure}

\begin{figure}
\includegraphics[width=9cm]{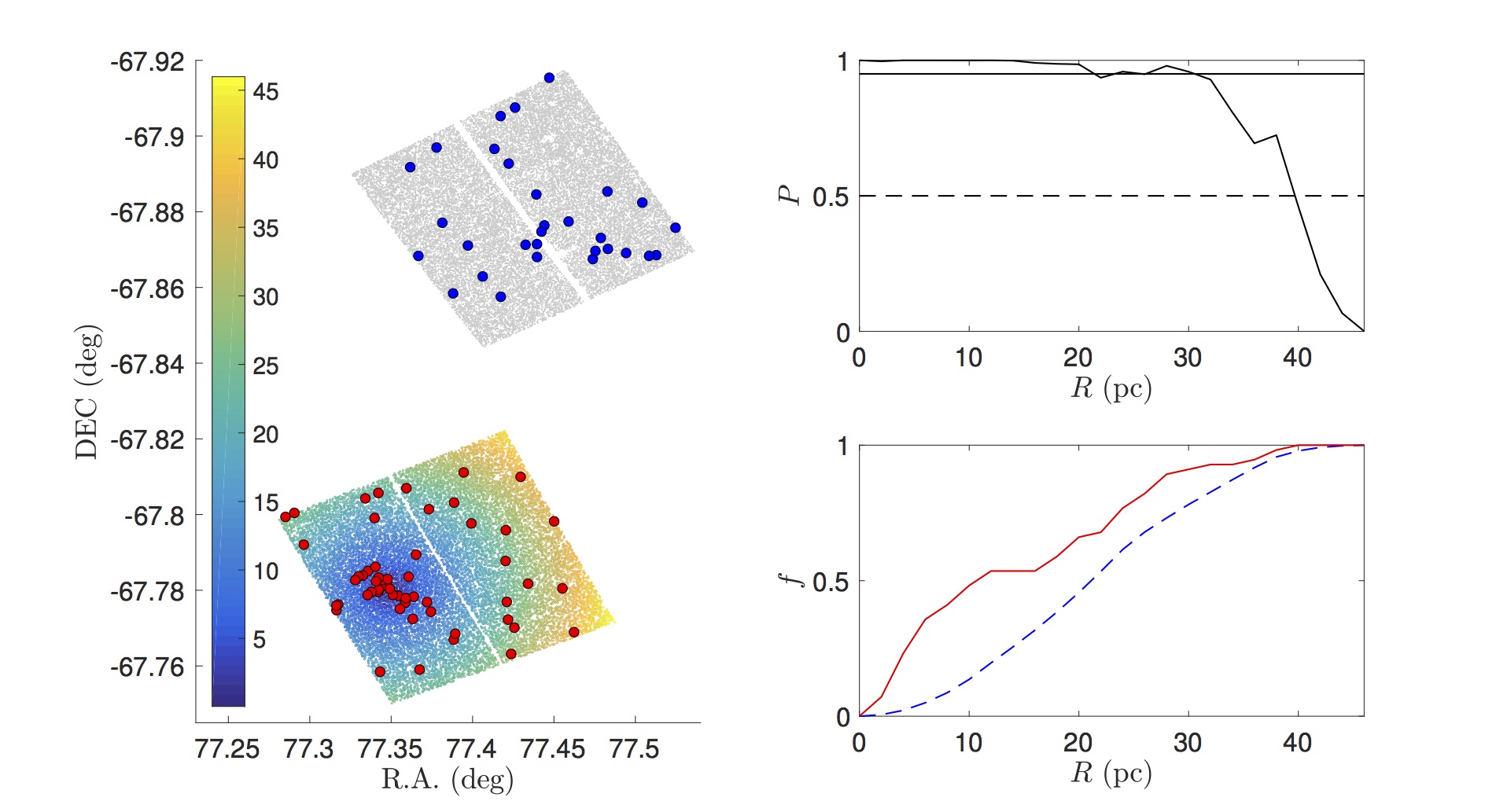}
\caption{As Fig. \ref{F6} but for NGC 1852.}
\label{F11}
\end{figure}

\begin{figure}
\includegraphics[width=9cm]{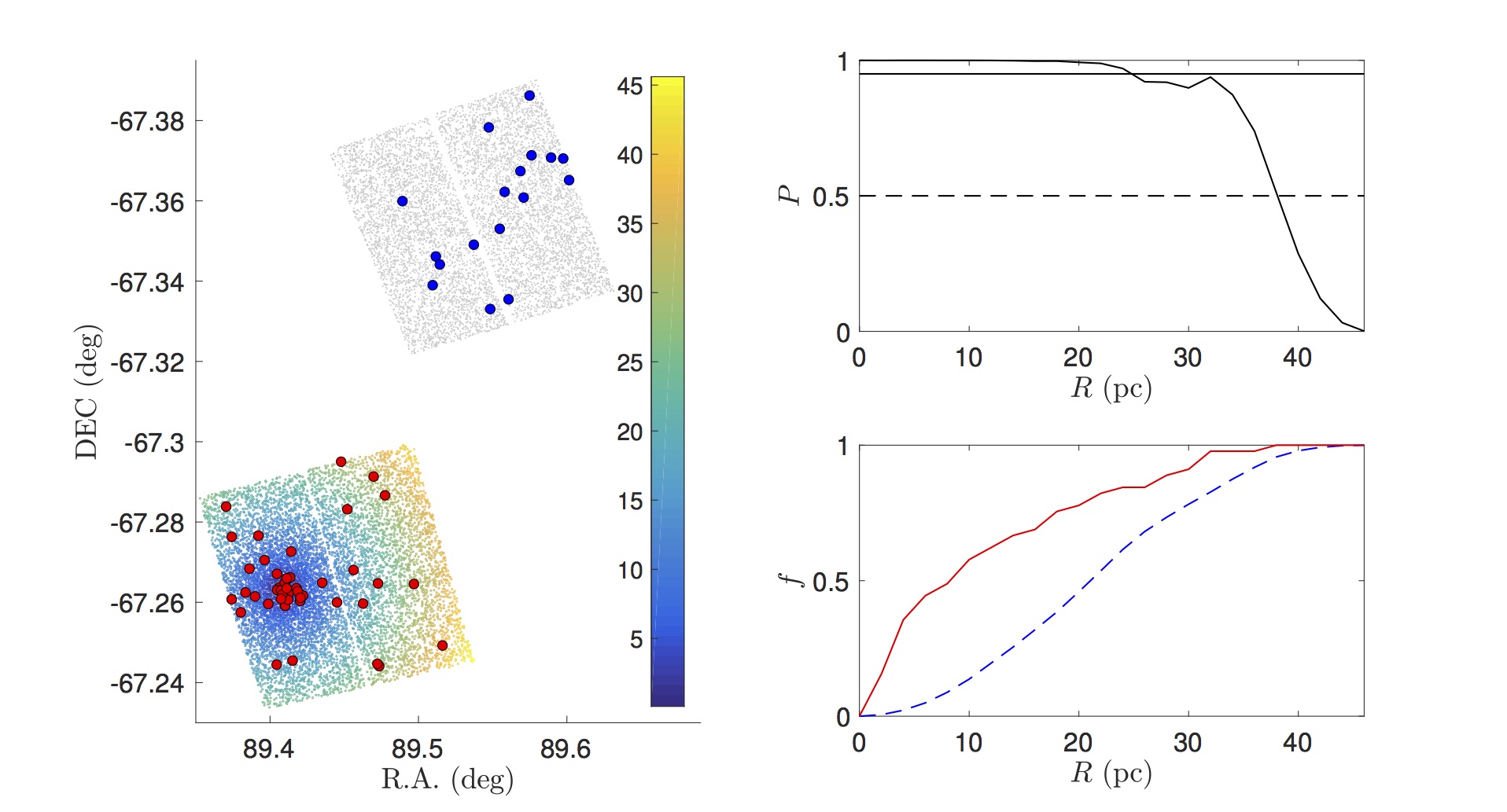}
\caption{As Fig. \ref{F6} but for NGC 2154.}
\label{F12}
\end{figure}

\begin{figure}
\includegraphics[width=9cm]{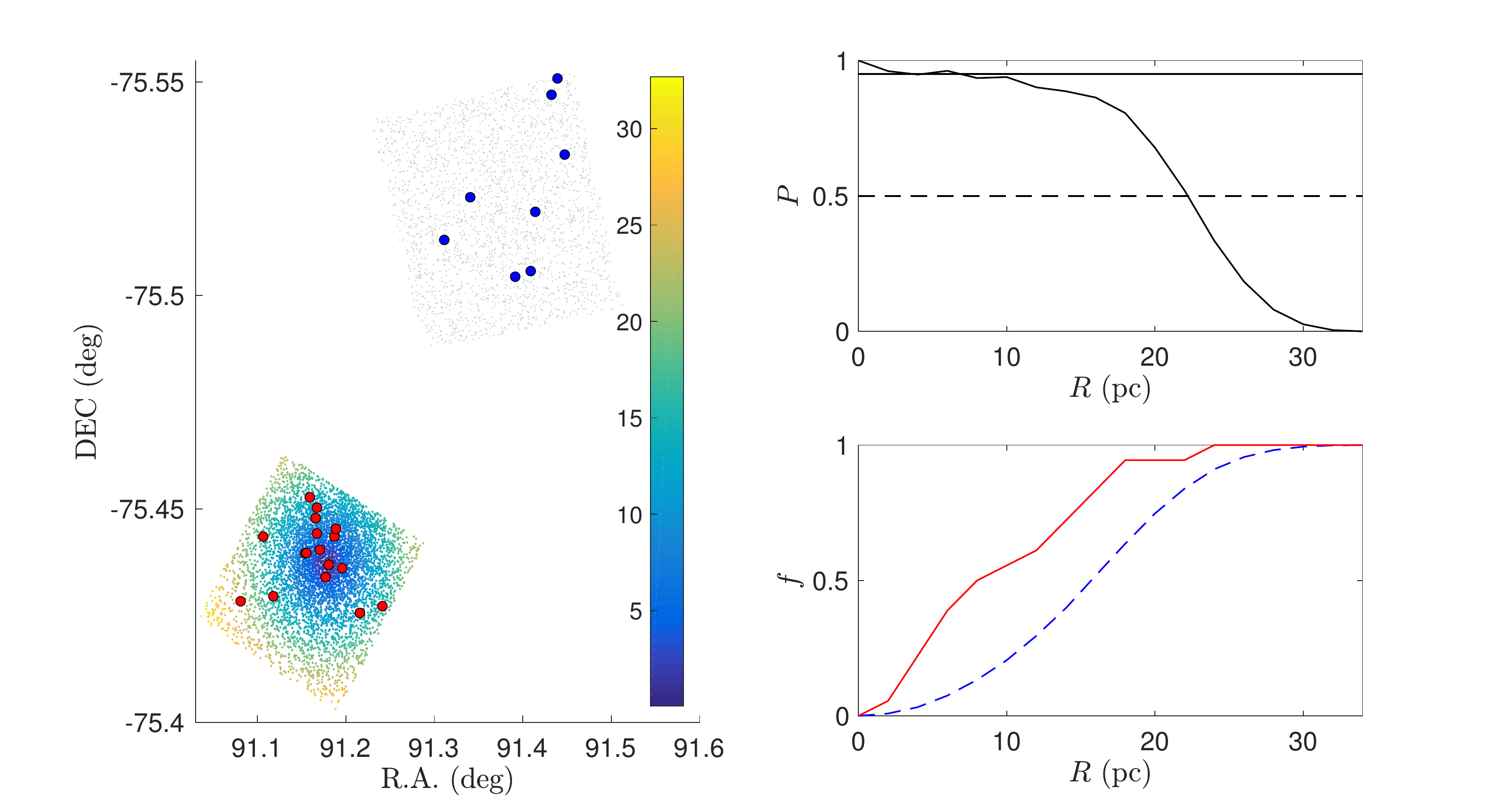}
\caption{As Fig. \ref{F6} but for NGC 2203.}
\label{F13}
\end{figure}

\begin{figure}
\includegraphics[width=9cm]{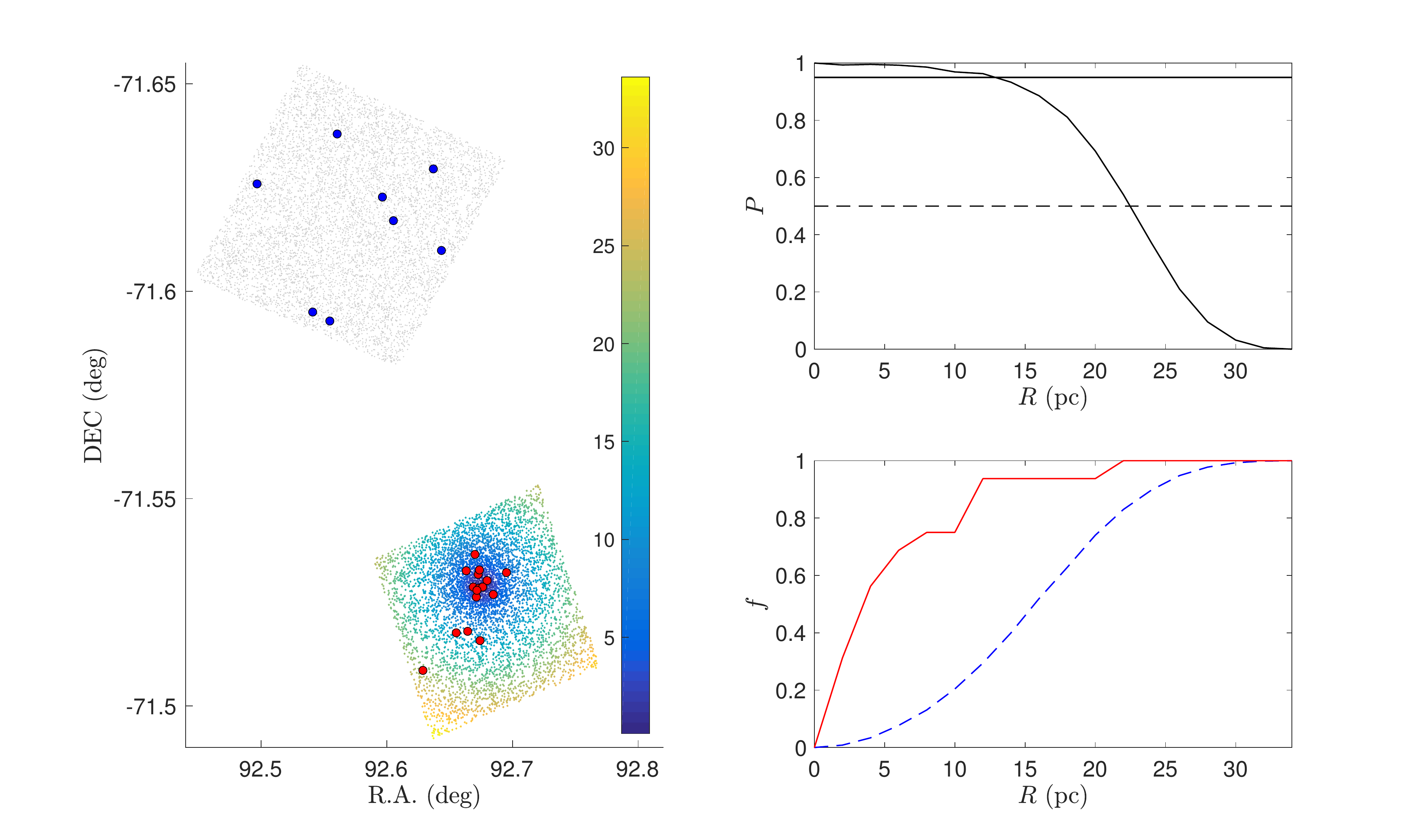}
\caption{As Fig. \ref{F6} but for NGC 2213.}
\label{F14}
\end{figure}

We present the spatial distributions of all selected bright evolved
stars and their field counterparts in Figs \ref{F6}--\ref{F14}
(left-hand panels). As shown in the right-hand panels of these
figures, all observed bright evolved stars are apparently concentrated
toward the cluster center. The spatial distributions of these bright
evolved stars are smoking guns, proving that most of these bright
evolved stars are cluster members rather than field stars.

We applied a Monte Carlo-based method to statistically quantify the
probability that the observed central concentrations of bright evolved
stars are simply derived from the underlying fields. To do so, for
each cluster we first calculated the observed central concentrations
at different radii. At a given radius $R$, the corresponding
concentration is defined as the number ratio of the stars within this
radius and all stars in the cluster region, $c=N(\leq{R})/N_{\rm
  tot}$. Then we randomly generated $N'$ field stars which were
homogeneously distributed in the cluster region, where $N'$ was
derived from a Gaussian distribution centered at $N_{\rm cont}$ (Table
\ref{T4}) with a Poisson-like uncertainty. Then, for these $N'$ field
stars, we calculated their corresponding concentration at radius $R$,
$c'=N'(\leq{R})/N'$. We repeated this procedure 10,000 times for each
radius and counted how many times we derived $c\leq{c'}$. This number,
divided by 10,000, was defined as the cluster membership probability
for stars inside this radius, $P=P(R)$. These cluster membership
probabilities, as a function of radius from the cluster center, are
presented in the top right-hand panels of Figs
\ref{F6}--\ref{F14}. Using the simulated field stars, we also
calculated their cumulative profiles. These field cumulative profiles
were used for comparison with those of the observed bright evolved
stars. These results are presented in the bottom right-hand panels of
Figs \ref{F6}--\ref{F14}.

Figures \ref{F6}--\ref{F14} have proved that field stars cannot
realistically reproduce highly concentrated spatial distributions such
as those of the observed bright evolved stars. If the observed
concentrations of bright evolved stars are simply produced by
homogeneously distributed field stars, our calculated cluster
membership probabilities should be $\leq50\%$ for all radii. In the
top right-hand panels of Figs \ref{F6}--\ref{F14}, we have highlighted
probability levels of 50\% (black dashed lines) and 95\% (black solid
lines). The latter indicates a strong concentration level. Clearly,
for most of our clusters the observed concentration levels are
significantly higher than what could be produced by field stars. This
is also illustrated by the difference between the cumulative profiles
of the observed eBSS candidates and the simulated field stars.

In summary, because (1) the numbers of field stars are significantly
lower than those of the observed bright evolved stars in the cluster
regions, and (2) the observed bright evolved stars are all spatially
concentrated in the central cluster regions, we conclude that most
detected bright evolved stars are most likely genuine cluster members
rather than field stars.

\section{Physical Implications}\label{S4}

\begin{figure*}
  \centering
  \includegraphics[width=2\columnwidth]{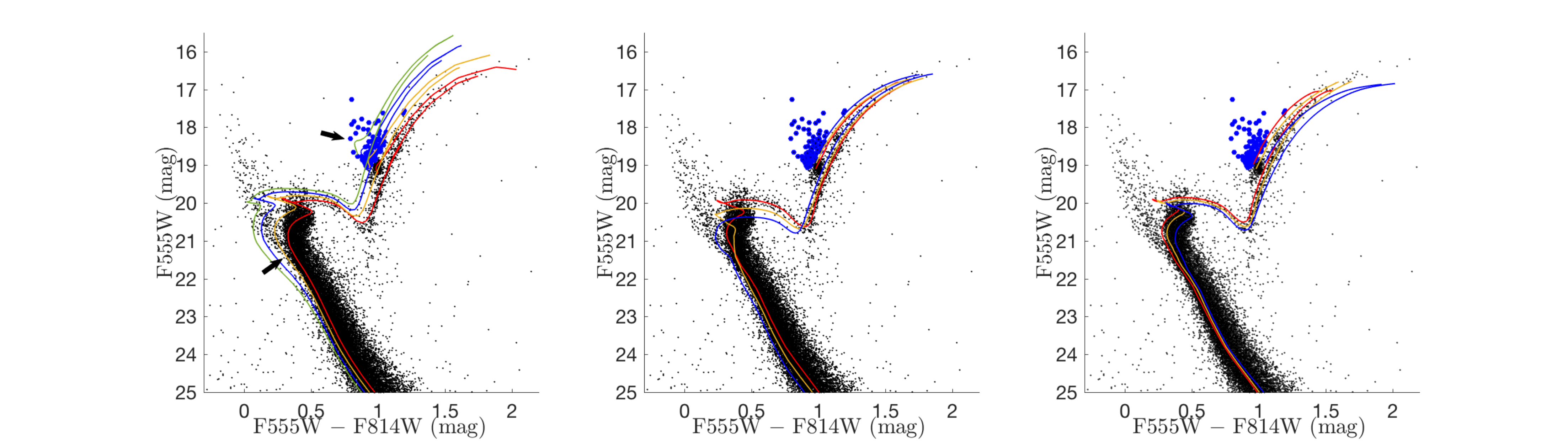}
  \caption{Isochrone fitting for NGC 1783. (left) All isochrones
    have fixed age but pertain to different metallicities. Red,
    orange, blue, and green isochrones correspond to
    $Z=0.006,0.004,0.002,0.001$, respectively. (middle) All isochrones
    have fixed age and metallicity but different helium
    abundances. Red, orange, and blue isochrones indicate $Y=0.254,
    0.33, 0.4$, respectively. (right) All isochrones have fixed age,
    metallicity, and helium abundance, but have different
    $\alpha$-element abundances, [$\alpha$/Fe]. Red, orange, and blue
    isochrones correspond to [$\alpha$/Fe] = $-0.2,0,0.2$ dex,
    respectively.}\label{F15}
\end{figure*}

We first examine if these bright evolved stars could be mimicked
  by a spread in their chemical abundances. In principle, spreads in
  the overall metallicity, helium, and $\alpha$-element abundances
  could cause spreads in both color and luminosity for a single-aged
  population of stars. Increasing the metallicity contributes to the
  prevailing stellar opacity, which would lead to an increase of the
  cooling efficiency. An enhanced metallicity would thus decrease the
  stellar surface temperature. In the meantime, the enhanced opacity
  would prevent the central energy from radiating to the stellar
  surface, making a metallicity-enhanced star look fainter than its
  lower-metallicity counterparts. Therefore, to reproduce an evolved
  stellar population (for a fixed age) that is obviously brighter and
  bluer than the normal giant stars, the stars must have a much lower
  metallicity. However, we found that for each of our clusters, even
  if we decreased their metallicity to $Z=0.001$ (which corresponds to
  almost 6\% of solar metallicity; note that all of our clusters have
  roughly half-solar metallicity, based on isochrone fitting), we
  could not reproduce the full set of observed bright evolved
  stars. In addition, such a large metallicity spread would produce a
  secondary MS that is bluer than the zero-age MS (ZAMS), which is not
  observed in our sample clusters. In the left-hand panel of
  Fig. \ref{F15}, we take the cluster NGC 1783 as an example. The
  adopted metallicities for the four isochrones are
  $Z=0.006,0.004,0.002,0.001$.

Another possible effect that might cause a spread of a
  single-aged stellar population in the CMD is related to the presence
  of a helium spread. Unlike metallicity differences, enhanced helium
  abundances would decrease the stellar radiative opacity, because the
  average opacity of helium is lower than that of hydrogen. On the
  other hand, an increased helium mass fraction would lead to a
  decrease of the hydrogen mass fraction, which would increase the
  mean molecular weight, thus leading to a stellar luminosity
  increase. The combination of these effects would make a
  helium-enhanced star hotter and brighter at each stage than its
  normal counterpart. However, such increased luminosities (and thus
  the hydrogen-burning efficiency) would shorten the stellar MS
  lifetimes. Therefore, a coeval stellar population containing
  different helium abundances would have a hotter and brighter MS and
  RGB, but a fainter MSTO and SGB.

To produce a population of brighter evolved stars with the same
  age as the normal stars, we have to increase their helium
  abundance. Because the helium abundance is not a free parameter in
  the PARSEC models, we used the Dartmouth stellar evolution database
  to generate a sample of isochrones with different helium abundances
  \citep{Dott07a,Dott08a}\footnote{\url{http://stellar.dartmouth.edu/models/index.html}}.
  The Dartmouth models offer three choices for the helium abundance,
  $Y=0.245+1.5Z$, 0.33, and 0.4. For all our clusters, the first
  choice we adopted was $Y=0.251,0.254,0.257$, corresponding to
  $Z=0.004,0.006,0.008$, respectively. However, the Dartmouth models
  do not include post-RGB phases in their isochrones. We therefore
  simply shifted the RGB to fit the AGB and the RC, since the AGB is
  roughly parallel to the RGB. This may lead to some additional
  uncertainties, but it should not significantly change our results,
  because an enhanced helium abundance should have the same effect on
  both RGB and AGB stars. Finally we confirm that even if we enhance
  the helium abundance to $Y=0.4$, this is still too small to
  reproduce the observed bright evolved stars. In addition, this will
  produce a broadened SGB in all clusters, while some of these
  clusters exhibit very tight SGBs instead
  \citep[e.g.,][]{Li14a,Bast15a}. We thus exclude the possibility that
  the observed bright evolved stars are helium-enhanced stars. In the
  middle panel of Fig. \ref{F15} we show isochrones defined by
  different helium abundances for NGC 1783 as an example.

The impact of enhanced $\alpha$ elements to isochrones has been
  studied by \cite{Sala93a}. The contribution of $\alpha$-element
  enhancements can be taken into account by simply rescaling standard
  models to the global metallicity, [M/H]:
\begin{equation}
[M/H]=[Fe/H]+\log(0.638f_{\alpha}+0.362),
\end{equation}
where $f_{\alpha}$ is the enhancement factor pertaining to the
$\alpha$ elements, $f_{\alpha}=10^{\rm [\alpha/Fe]}$. For the solar
$\alpha$ abundance $[\alpha/Fe]=0$, this relation obviously
yields [M/H] = [Fe/H]. Therefore, the effect of a spread in $\alpha$
elements is similar to that of a metallicity spread. To produce a
population of brighter evolved stars, we need to decrease the $\alpha$
abundances. Again, we employed a sample of isochrones with different
$\alpha$-element abundances, down to [$\alpha$/Fe]=$-0.2$\footnote{The
  minimum value in the Dartmouth stellar evolution database.}, but
with fixed [Fe/H], $Y$, and age. We found that the effect of an
$\alpha$-element spread is negligible compared to the color spread of
these bright evolved stars (see the right-hand panel of Fig. \ref{F15}
for NGC 1783 as an example).

In summary, we have not found any evidence to suggest that these
  bright evolved stars can be fully reproduced by an internal chemical
  spread. The only viable explanation is that they are much younger
  than the clusters' bulk population stars. Here the question is
  whether they are genuine young stars formed through a continuous
  star-formation process or if they have a dynamical origin, e.g., as
  for BSSs.

All our clusters harbor extended MSTO (eMSTO) regions
  \citep[e.g.,][]{Milo09a}. Initially, the most straightforward
  explanation for the eMSTO region of individual clusters was thought
  that these clusters could have experienced an extended
  star-formation history lasting several hundred million years
  \citep[e.g.,][]{Goud11a,Goud14a}. However, discussions in the
  literature soon pivoted to an explanation based on the presence of a
  population of rapidly rotating stars \citep{Bast09a,Bran15a},
  because the extended star formation history scenario suffers from
  many difficulties when dealing with details in other CMD regions
  \citep{Li14a,Li16a} and when considering global features pertaining
  to the inferred age spread and the isochronal age of the clusters
  \citep{Nied15a}. However, some recent studies have suggested that
  some clusters may still require a genuine age spread to reproduce
  their very wide MSTO regions \cite{Goud17a,Piat17a}. This discussion
  continues unabated.
 
If the scenario proposed by \cite{Goud11a} is on the right track,
  this may indicate that the observed bright evolved stars in our
  clusters are genuine stars formed through a continuous star-forming
  process. However, all of the bright evolved stars are located beyond
  an isochrone that is at least {\color{red} 600 Myr} younger than the bulk
  population stars (see Table \ref{T3}), where \cite{Goud11a} have
  determined that the star-formation histories of star clusters such
  as those observed in the Magellanic Clouds can only last for up to
  $\sim$500 Myr. Therefore, an extended period of star formation seems
  unable to explain these observed bright evolved stars. In
  particular, if the presence of these very bright stars would
  indicate an even longer star-formation history in these clusters, we
  would expect to see some pre-main-sequence (PMS) stars with ages of
  200--500 Myr. Such PMS stars may exhibit an H${\alpha}$ excess,
  which can be easily detected in the {\sl HST} UVIS/WFC3 F656N
  passband. Such a survey has recently been carried out by
  \cite{Milo18a}, aiming to search for Be stars in young LMC
  clusters. No obvious PMS features were detected in their clusters,
  which thus indirectly supports the notion that the star-forming
  process in many LMC clusters cannot last so long.

Mergers of two clusters with different ages \citep{Hong17a}, or of a
cluster and a giant molecular cloud \citep[GMC;][]{Bekk09a}, may
produce a stellar system composed of multiply aged stellar
populations. LMC and SMC systems may have experienced close encounters
some 100 Myr and 1--2 Gyr ago \citep{Gard96a}. If such an LMC--SMC
tidal interaction would trigger frequent interactions between clusters
or between clusters and GMCs, this may produce large numbers of
multiply aged stellar population systems exhibiting similar age
differences. However, this scenario cannot explain why the bright
evolved stars in our clusters all represent a small number fraction
compared with the bulk population stars, since a merger event can
involve two stellar systems with various mass ratios.

The most straightforward explanation is that these bright evolved
stars are eBSSs. Since the mass limit for BSSs is twice the mass of
MSTO stars, their evolved counterparts must have the same mass limit
as well. The upper mass limit for our young evolved stars is simply
defined by the adopted youngest isochrone. We use the mass of the
bottom of the RGB of the youngest isochrone as the typical maximum
mass of our young evolved stars.  Correspondingly, the mass of the
bottom of the RGB of the old isochrone is the typical mass of normal
evolved stars. We than check if the former mass would be higher than
twice the latter. In Table \ref{T3}, we list the masses for the
  bottom of the RGB corresponding to the bulk population and to two
  young isochrones (columns 5--7), and the mass ratio of the upper
  mass limit to twice the mass for the bottom of the RGB of the bulk
  stellar population (column 8). We also list the age for a stellar
  population with exactly twice the mass for the bulk population
  stars. We found that, except for NGC 1783, all clusters should have
  bright evolved stars that are less massive than twice the mass of
  their normal evolved counterparts. It is likely that these observed
  bright evolved stars are eBSSs formed through interactions between
  first-generation stars. For NGC 1783, there is only one star
   slightly more massive than twice the mass of the normal
  evolved stars, as indicated by the arrow in Fig. \ref{F3}. Given
  that NGC 1783 contains the largest sample of eBSS candidates (100),
  one exception may simply be caused by fitting uncertainties and does
  not change our conclusions.

We found that all of our clusters would have bright evolved stars that
are less massive than twice the mass of their normal evolved
counterparts. It is likely that these observed bright evolved stars
are eBSSs formed through interactions between first-generation stars.

To further explore their origin, we calculated the collisional rate
for the core regions of our clusters \citep{Davi04a},
\begin{equation}
N_{\rm col}=0.03225\frac{f^2_{\rm mms}N_{\rm c}n_{\rm c,5}r_{\rm
    col}m_{\rm BSS}}{V_{\rm rel}},
\end{equation}
where $N_{\rm col}$ represents the number of expected collisional BSSs
that would be produced over the last 1 Gyr, $f_{\rm mms}$ is the
fraction of massive MS stars in the core region (where `massive MS
star' means they are sufficiently massive to form a collisional BSS),
and $f_{\rm mms}=0.25$ is commonly adopted \citep{Davi04a}. $N_{\rm
  c}$ is the number of stars located in the core region. We evaluated
this number as follows. We counted the number for MS stars with
absolute F814W magnitudes of $\sim$2--3 mag (equal to a stellar mass
of $\sim$1.2--1.5 $M_{\odot}$). We assumed that these stars follow a
Kroupa-like mass function. Then, we evaluated the total number of
stars by extrapolating this mass function down to $0.08 M_{\odot}$. We
selected stars in this magnitude range, because they will have a high
completeness and follow a simple mass--luminosity
relationship. $n_{c,5}$ is the stellar number density for the core
region in units of $10^5$ pc$^{-3}$, which is the ratio of the number
of stars contained in the core region and the core volume. $r_{\rm
  col}$ represents the minimum separation of two colliding stars in
units of the solar radius, $R_{\odot}$. We assume that it is twice the
stellar radius of the BSSs. We assume that the average mass of BSSs is
equal to that of our observed young evolved stars, as indicated by the
best-fitting isochrones. Their radii were evaluated through the
empirical formula introduced by \citep{Demi91a}. $V_{\rm rel}$ is the
relative incoming velocity of binaries,
\begin{equation}
V_{\rm rel}=\sqrt{2}\sigma=\sqrt{\frac{4GM_{\rm c}}{r_{\rm c}}},
\end{equation}
where $\sigma$ is the velocity dispersion for all stars in the core
region. $M_{\rm c}$ is the stellar mass of the cluster core, since we
have already evaluated the total number of stars in the core region
following a Kroupa mass function, $M_{\rm c}$ is simply the sum of the
masses of these stars. However, it is possible that we have
overestimated the number of stars in the core, since numerous low-mass
stars must have evaporated from the central cluster region owing to
two-body interactions \citep[e.g.,][]{grijs02a}.

The resulting stellar collisional rate ($N_{\rm coll}$, per Gyr), as
well as the derived global parameters ($N_{\rm c}$, $n_{\rm c,5}$,
$\sigma$, $M_{\rm c}$) for our clusters, are summarized in Table
\ref{T5}.

\begin{table*}
  \begin{center}
\caption{Derived physical parameters.}\label{T5}
  \begin{tabular}{c c c c c c c}\hline
    Cluster      &  $N_{\rm c}$ & $n_{\rm c,5}$ ($10^5$ pc$^{-3}$)& $\sigma$ (km s$^{-1}$) & $M_{\rm c}$ (M$_{\odot}$) & $N_{\rm coll}$ & $\rho$ (M$_{\odot}$ pc$^{-3}$) \\
        	&	(1)	&	(2)	&	(3)	&	(4)	&	(5)	&  (6)\\\hline
    NGC 1644 & 4.02$\times10^3$ & 0.0078 & 3.28 & 1.33$\times10^3$ & 0.121 & 282.69\\
    NGC 1651 & 1.74$\times10^4$ & 0.0009 & 3.68 & 5.64$\times10^3$ & 0.059 & 29.10\\
    NGC 1783 & 6.95$\times10^4$ & 0.0016 & 6.53 & 2.32$\times10^4$ & 0.293 & 54.03\\
    NGC 1806 & 4.82$\times10^4$ & 0.0030 & 6.23 & 1.53$\times10^4$ & 0.359 & 93.76\\
    NGC 1846 & 5.44$\times10^4$ & 0.0007 & 5.14 & 1.78$\times10^4$ & 0.120 & 21.67\\
    NGC 1852 & 1.98$\times10^4$ & 0.0010 & 3.98 & 6.78$\times10^3$ & 0.076 & 8.32\\
    NGC 2154 & 1.41$\times10^4$ & 0.0019 & 3.90 & 4.63$\times10^3$ & 0.112 & 62.17\\
    NGC 2203 & 1.30$\times10^4$ & 0.0007 & 3.17 & 4.24$\times10^3$ & 0.041 & 21.34\\
    NGC 2213 & 3.26$\times10^3$ & 0.0024 & 2.48 & 1.06$\times10^3$ & 0.056 & 78.06\\\hline
  \end{tabular} 
  \end{center} 
  (1) Expected number of stars in the core region. (2) Central number
  density (in units of $10^5$ pc$^{-3}$). (3) Central velocity
  dispersion. (4) Core mass. (5) Expected number of collisional BSSs
  formed during the last 1 Gyr.
\end{table*} 

Our calculation shows that it is unlikely that these clusters could
produce sufficient numbers of collisional BSSs at their relevant ages,
at least for single--single collisional BSSs. However, binary-mediated
stellar collisions may dominate the formation of collisional
BSSs. \cite{Chat13a} modeled 128 GGCs with various properties to
investigate the dominant formation channels of BSSs in these
clusters. They find that for central stellar mass density higher than
$\sim10^3$ M$_{\odot}$ pc$^{-3}$, binary-mediated stellar collisions
will make a major contribution for BSS formation. In Table \ref{T5} we
have also listed the mass densities in the core regions of our
clusters,
\begin{equation}
\rho_{\rm c}=\frac{3M_{\rm c}}{4\pi{r_{\rm c}^3}}
\end{equation}
As we can see, even the densest cluster, NGC 1644, does not have a
core density that reaches the density threshold of $\sim10^3
M_{\odot}$ pc$^{-3}$. Therefore, binary-mediated stellar collisions
are also unlikely major formation channels for these eBSSs in our
clusters. If these young evolved stars are eBSSs, they are therefore
all most likely produced through binary interactions.

It is also possible that some of these bright evolved stars are not
massive stars, but that they are instead unresolved binary systems,
which may appear brighter than single stars in the CMD. To explore
this possibility, for each cluster, we calculated the loci of the
binary sequences for different mass ratios, $q=M_2/M_1$. Here, $M_1$
and $M_2$ are masses of the primary and secondary components,
respectively. The loci of these binary sequences are calculated as
follows.
\begin{itemize}
\item[i.]{We selected the best-fitting isochrone to the bulk
  population as our baseline.}
\item[ii.]{For each point of the baseline, we calculated the secondary
  star's mass for a fixed mass ratio ($M_2=qM_1$).}
\item[iii.]{Calculate the magnitude in each passband for the secondary
  star through interpolation, based on the selected isochrone.}
\item[iv.]{Calculate the resulting magnitude,
\begin{equation}
m = -2.5\log{(10^{-0.4m_1}+10^{-0.4m_2})},
\end{equation}
where $m$, $m_1$, and $m_2$ are magnitudes of the binary system and
its primary and secondary components, respectively.}
\item[v.]{Connect all points for the resulting magnitudes calculated
  in step (iv), which is the locus of a binary sequence with a mass
  ratio $q$.}
\end{itemize}

\begin{figure}
  \centering
  \includegraphics[width=\columnwidth]{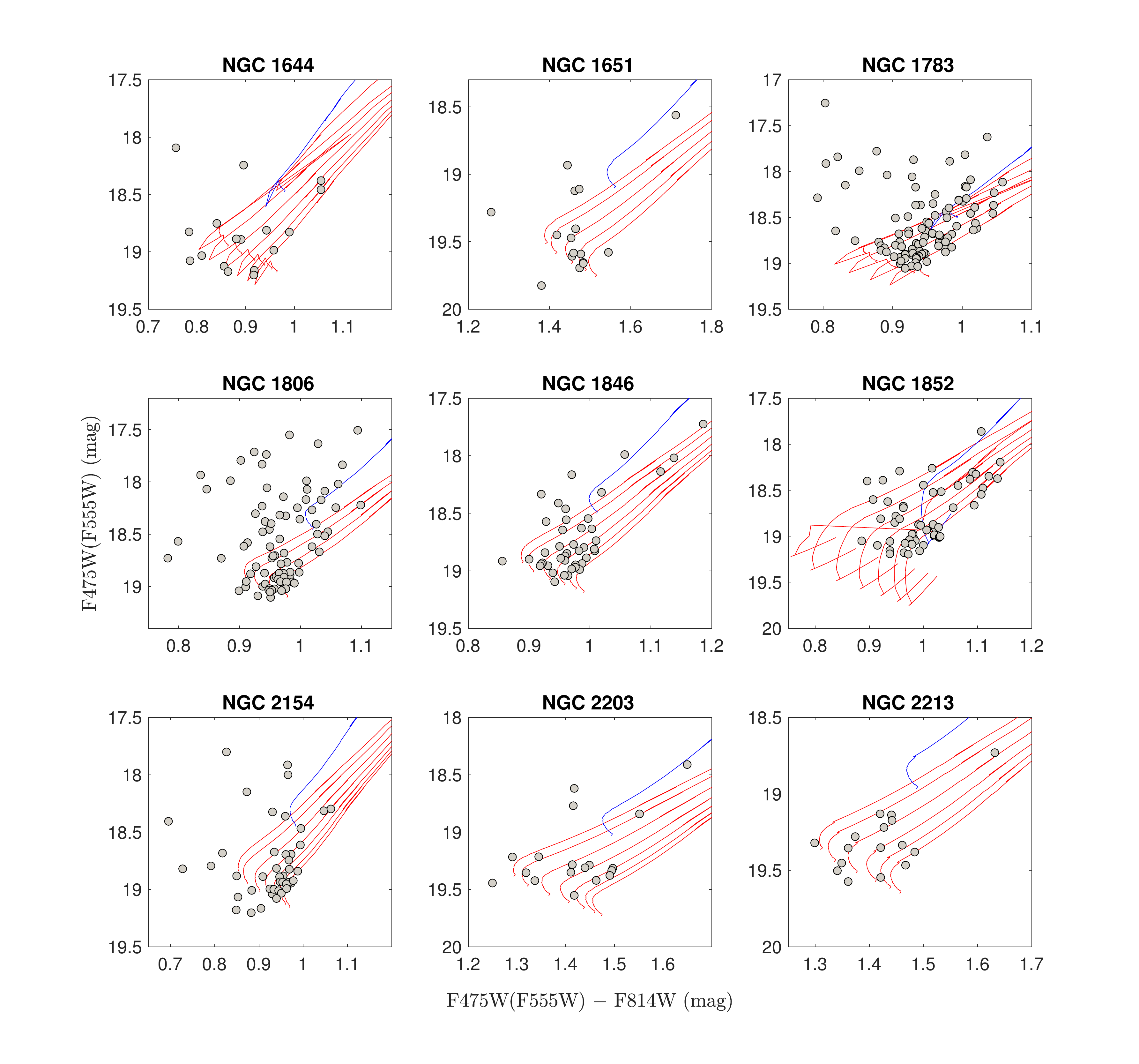}
  \caption{Observed bright evolved stars in our clusters as well as
    binaries sequences for different mass ratios. The blue solid lines
    represent the equal-mass binary sequences. The red solid lines are
    binary sequences for mass ratios of less than one, from bottom to
    top the mass ratio increases.}\label{F16}
\end{figure}

In Fig. \ref{F16}, we present the calculated binary sequences,
compared with the observed young evolved stars in the CMDs. It is
clear that only high mass-ratio binaries (with $q\geq0.65$--$0.75$)
will have their unresolved photometric magnitudes significantly
different from the loci of the single stars. This is because our stars
are so bright that only a secondary star with comparable luminosity
would significantly affect the photometry. The secondary star must be
very massive, probably a giant star. This is because compared with a
giant star, a low-mass MS component is too faint to affect the
photometry. In Fig. \ref{F16}, the red curves from right to left are
binary loci with mass ratios of $q=0.65$ to 0.95. These loci are all
bluer than the normal RGB because these binaries all contain a
secondary star that is bluer than the primary, evolved star. Once the
mass ratio reaches unity, $q=1$, the binary locus moves back to the
red side, because in that case all binaries simply contain two
equal-mass giant stars. It is parallel to the bulk stellar population
but it is $-0.752$ mag brighter.

As shown in Fig. \ref{F16}, many bright evolved stars observed in our
clusters are located in the region covered by binary sequences for
different mass ratios. Therefore, it is possible that the bright
evolved stars are not massive stars but unresolved high mass-ratio
binary systems. In particular for NGC 1852, almost all of its bright
evolved stars are located in the region of high mass-ratio
binaries. However, unresolved binaries only cover a very small region
in the CMD. They are able to explain some stars which are very close
to the normal stellar population. To explain stars that are very blue
and bright, like those in NGC 1783, NGC 1806, or NGC 2154, a
significant population of massive stars is still required.

\begin{figure}
  \centering
  \includegraphics[width=\columnwidth]{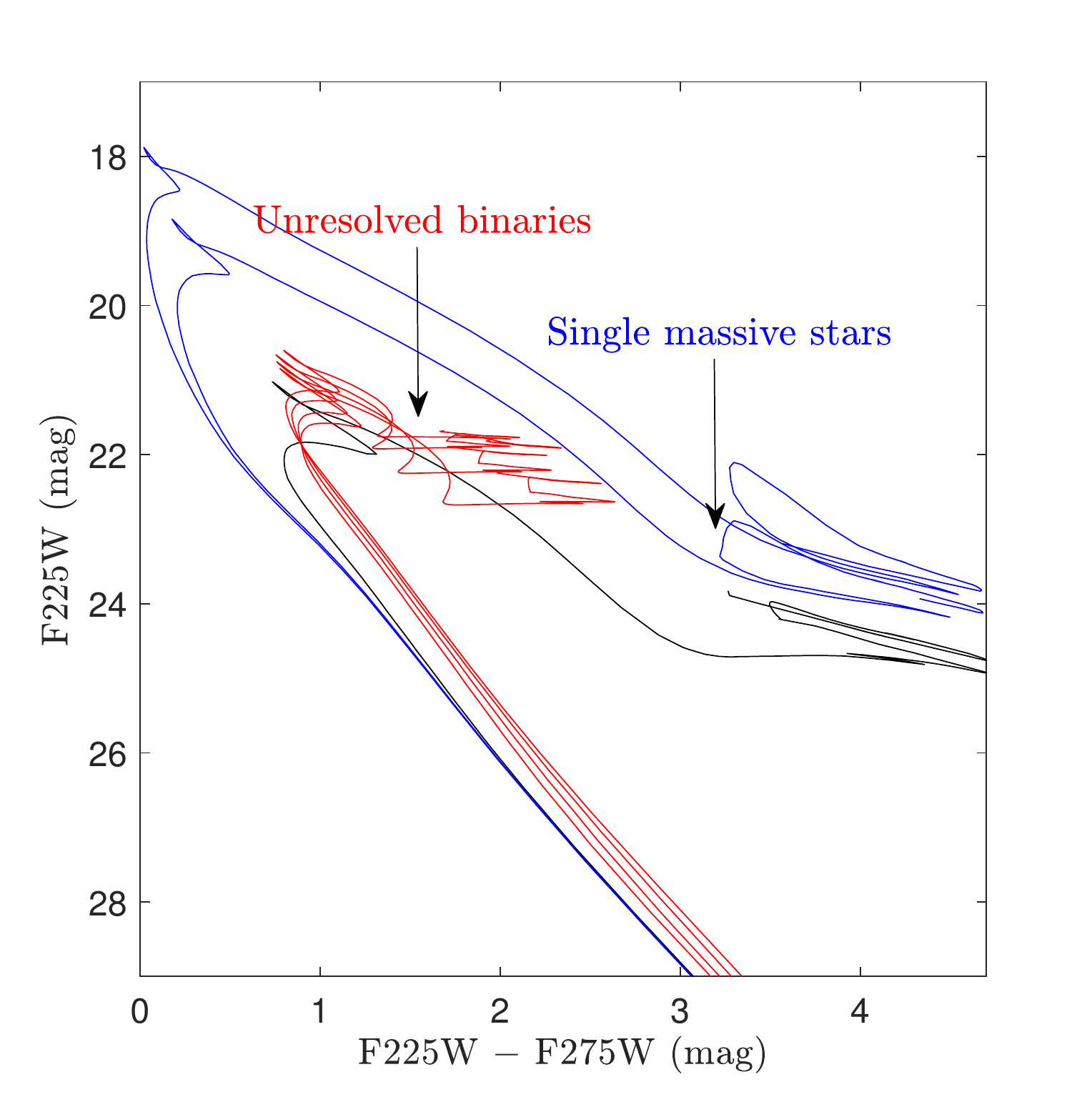}
  \caption{Illustration of how far-UV CMDs (based on the F225W and
    F275W passbands) could allow us to distinguish unresolved binaries
    from single massive evolved stars, taking NGC 2203 as an
    example. The red solid curves represent unresolved binaries with
    different mass ratios. The blue and black curves are young and old
    isochrones, respectively, adopted for the cluster. All isochrones
    were generated by the PARSEC stellar evolution model
    \citep{Bres12a}.}\label{F17}
\end{figure}

To separate massive single stars and unresolved binaries, observations
in far-ultraviolet (far-UV) passbands would provide useful
information, such as in the {\sl HST} UVIS/WFC3 F225W and F275W
bands. In the far-UV, the less-evolved stars are even brighter than
the more evolved stars, since the latter would move to the red. For
unresolved binaries composed of two first-generation stars, the total
flux of these binary systems might be dominated by the less-evolved
components. The resulting photometry for these binary systems will be
close to the MS in the CMD. On the other hand, genuine massive evolved
stars will appear much redder and brighter than unresolved
binaries. In Fig. \ref{F17}, taking NGC 2203 as an example, we show
the adopted young isochrones, as well as the predicted loci of binary
sequences with different mass ratios, in the F225W--F275W CMD. The
positions of the unresolved binaries and genuine young evolved stars
are indeed well separated in the CMD.

In summary, the detected bright evolved stars in our clusters are most
likely eBSSs, it is also possible that some are simply unresolved
binaries.

If these bright evolved stars are indeed eBSSs, it seems that their
numbers are large, even comparable to the numbers of normal BSSs
observed in old GCs \citep[e.g.,][]{Leig13a}. However, this does not
mean that larger numbers of normal BSSs hide in the MS. Because BSSs
are formed through stellar dynamics rather than the collapse of GMCs,
it is not necessary that they should form according to a Kroupa-like
mass function \citep{Krou01a}. In fact, early-type stars usually have
higher binary fractions \cite{Duch13a}, and the overall binary
fraction in a dense cluster will continue to decrease over time
because of dynamical destruction \citep{Ivan05a}. Given that most of
the progenitors of our eBSSs should be B-type stars, we are not
surprised about the large numbers of eBSSs in these clusters. Indeed,
\cite{Hypk17a} have simulated the number evolution of BSSs in dense
clusters for different initial conditions. They found that at ages of
4--5 Gyr, the number of BSSs is expected to reach a peak. For clusters
aged 1--2 Gyr, the expected number of BSSs is comparable to that in
old GCs ($\geq$10 Gyr).

\section{Conclusion}\label{S5}

Based on high-precision multi-band {\sl HST} photometry, we analyzed
the CMDs of nine young GCs in the LMC. We found that all these
clusters harbor samples of bright evolved stars. After having ruled
out the effects of field contamination and photometric artifacts, we
conclude that these bright evolved stars are likely genuine cluster
members. The main results can be summarized as follows.

\begin{itemize}
\item The combination of photometric scatter, internal distance
  spread, and differential reddening cannot explain the large
  color--magnitude dispersion of these bright evolved stars. These
  stars can be well described by isochrones with younger ages, 
    but not by isochrones for different metallicities, helium, or
    $\alpha$-element abundances, which means that they are more
  massive than normal stars, described by an older isochrone.
\item Statistical analysis shows that the reference field stars cannot
  fully explain nor generate these bright evolved stars. Compared with
  field stars with the same color--magnitude distributions, these
  bright evolved stars are overdense in the cluster region. In
  addition, their spatial distributions exhibit clear central
  concentrations, which cannot be explained by a homogeneous field.
\item Unresolved binaries can partially reproduce the color--magnitude
  distributions of these bright evolved stars. However, such binaries
  can only cover a very compact region in the CMD. For some stars that
  are extremely blue and bright, we still require a significant
  fraction of massive population stars.
\item If we assume that all these bright evolved stars are single
  stars, their masses would not exceed twice the mass of normal
  evolved stars. We suggest that this may indicate that most of these
  bright evolved stars are evolved products of their first-generation
  stars, that is, they are eBSSs. Our dynamical calculations show that
  none of our clusters can produce sufficient numbers of collisional
  BSSs over their lifetimes. Therefore, the only viable explanation
  for these eBSSs is that they formed through binary interactions.

\end{itemize}

\
\section{Acknowledgements}
We thank the anonymous referee for valuable comments. C. L. is
supported by the Macquarie Research Fellowship Scheme. This work was
supported by the National Key Research and Development Program of
China through grant 2017YFA0402702. L. D. and R. d. G. also
acknowledge research support from the National Natural Science
Foundation of China (grants U1631102 and 11373010). J. H. acknowledges
support from the China Postdoctoral Science Foundation, Grant
No. 2017M610694. Parts of this research were conducted with the
support of the Australian Research Council Centre of Excellence for
All Sky Astrophysics in 3 Dimensions (ASTRO 3D), through project
number CE170100013.

\vspace{5mm}
\facilities{{\sl Hubble Space Telescope} (WFC3/UVIS and ACS/WFC)}


\software{
{\sc dolphot2.0} \citep{Dolp16a} 
          }

\end{document}